\newif\ifpr

\ifpr
\documentclass[amsmath, amssymb, aps, prl, superscriptaddress, twocolumn, floatfix, 10pt]{revtex4-2}
\newcommand{\arxivPR}[2]{#2}
\else
\documentclass[amsmath, amssymb, aps, prl, superscriptaddress, twocolumn, floatfix, 10pt, nofootinbib]{revtex4-1}
\newcommand{\arxivPR}[2]{#1}
\newcommand{\secref}[1]{Sec.\,\ref{#1}}
\fi
\usepackage[utf8]{inputenc}

\usepackage{gensymb,amssymb,graphicx,color}
\usepackage{hyperref}
\usepackage[normalem]{ulem}
\usepackage[caption=false]{subfig} 

\newcommand{\arxiv}[1]{\arxivPR{#1}{}}

\newcommand{\refcite}[1]{Ref.\,\cite{#1}}
\newcommand{\refscite}[1]{Refs.\,\cite{#1}}
\newcommand{\eqnref}[1]{Eq.\,\eqref{#1}}
\newcommand{\eqsref}[1]{Eqs.\,\eqref{#1}}
\newcommand{\figref}[1]{Fig.\,\ref{#1}}
\newcommand{\appref}[2]{\arxivPR{Appendix~\ref{#1}}{SM~#2~\cite{supplementary}}}
\newcommand{\apprefFst}[2]{\arxivPR{Appendix~\ref{#1}}{Supplemental Material (SM) #2~\cite{supplementary}}}
\newcommand{\mysec}[1]{\arxivPR{\section{#1}}{\textbf{\textit{#1}} ---}}
\newcommand{\mysubsec}[1]{\arxivPR{\subsection{#1}}\textit{#1}:}

\newcommand{\ii}{\mathrm{i}}
\newcommand{\dd}{d}

\newcommand{\ket}[1]{|#1 \rangle}

\newcommand{\nn}{\nonumber}

\newcommand{\ZZ}{\mathbb{Z}}
\newcommand{\NN}{\mathbb{N}}
\newcommand{\RR}{\mathbb{R}}
\newcommand{\M}{\mathcal{M}}

\newcommand{\nf}{{n_\text{f}}}
\newcommand{\trm}{\mathrm{t}}
\newcommand{\xrm}{\mathrm{x}}
\newcommand{\yrm}{\mathrm{y}}
\newcommand{\zrm}{\mathrm{z}}
\newcommand{\lt}{{l_\trm}}
\newcommand{\lx}{{l_\xrm}}
\newcommand{\ly}{{l_\yrm}}
\newcommand{\lz}{{l_\zrm}}

\definecolor{kspink}{RGB}{200,0,200}

\hypersetup{
  colorlinks = true,
  urlcolor = blue,
  pdfauthor = {Kevin Slagle},
  pdftitle = {Slagle - 2020 - Foliated Quantum Field Theory of Fracton Order}
}

\begin{document}

\title{Foliated Quantum Field Theory of Fracton Order}

\author{Kevin Slagle}
\affiliation{Department of Physics and Institute for Quantum Information and Matter, California Institute of Technology, Pasadena, California 91125, USA}
\affiliation{Walter Burke Institute for Theoretical Physics,
California Institute of Technology, Pasadena, California 91125, USA}


\begin{abstract}
We introduce a new kind of \emph{foliated quantum field theory} (FQFT) of gapped fracton orders in the continuum.
FQFT is defined on a manifold with a layered structure given by one or more foliations,
  which each decompose spacetime into a stack of layers.
FQFT involves a new kind of gauge field, a \emph{foliated gauge field},
  which behaves similar to a collection of independent gauge fields on this stack of layers.
Gauge invariant operators (and their analogous particle mobilities) are constrained to the intersection of one or more layers from different foliations.
The level coefficients are quantized and exhibit a duality that spatially transforms the coefficients.
This duality occurs because the FQFT is a \emph{foliated fracton order}.
That is, the duality can decouple 2+1D gauge theories from the FQFT through a process we dub \emph{exfoliation}.
\end{abstract}

\maketitle

Fracton topological order
  \cite{FractonRev1,FractonRev2,VijayFracton,VijayXcube,HaahCode,BravyiFracton}
  is a phase of matter that exhibits particles with mobility constraints.
Such particles include fractons, lineons, and planons,
  which are energetically constrained to 0-dimensional, 1-dimensional, and 2-dimensional spatial submanifolds
  when isolated from other excitations.
Fracton research has been motivated as a means for more robust quantum information storage
  \cite{HaahCode,HaahMemory,BrownQuantumMemory},
  novel dynamics
  \cite{ChamonGlassy,PremGlassy,PaiCircuits,GromovHydrodynamics,PaiScars,PremLSM,HughesLSM,ScheurerPT,PollmannMoment,PengSuperfluid},
  toy models for holography \cite{YanHolography,YanHolography2},
  exotic materials and fluids 
  \cite{YanBreathing,PretkoElasticity,PretkoDuality,HsiehChains,FujiLayer,Slagle2spin,PretkoPinch,GromovDuality,GromovVortices,SondhiPlaquettes,CenkePlaquettes,SousPolarons,YizhiLego},
  and connections to quantum gravity \cite{PretkoGravity}.

\begin{figure}
  \includegraphics[width=\columnwidth]{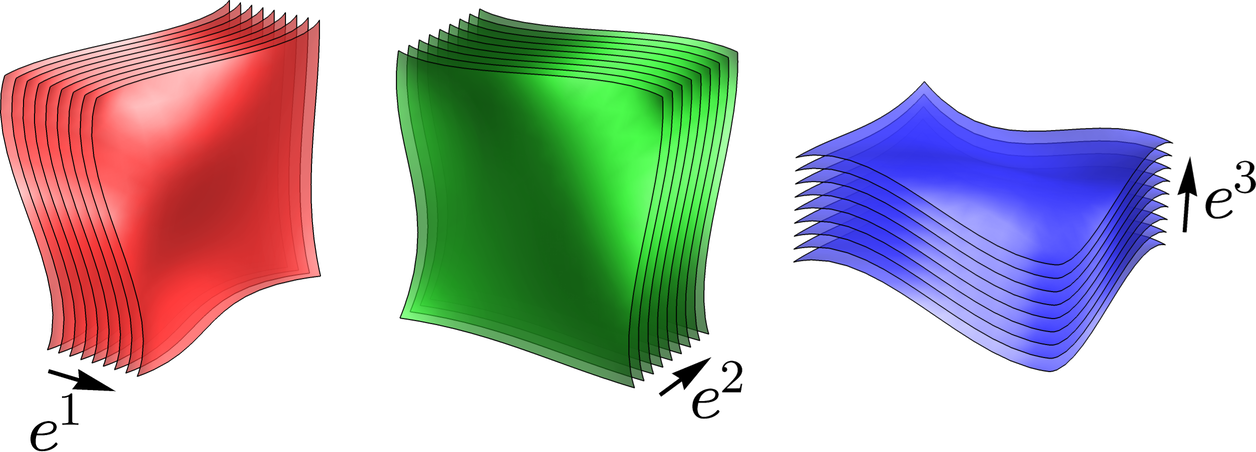}
  \caption{%
    A depiction of some leaves (colored surfaces) for three different foliations.
    A foliation consists of an infinite number of infinitesimally-spaced layers, which are called leaves.
  }\label{fig:foliation}
\end{figure}

In this work, we focus on gapped\footnote{%
  We will not study the gapless $U(1)$ fracton models \cite{PretkoU1,RasmussenU1,PretkoEM,RadzihovskyVector,SeibergU1,SeibergSymmetry,PretkoTemperature,PolyShift},
    which are analogous to $U(1)$ Maxwell gauge theory.
  Gapped fracton models are analogous to $Z_N$ gauge theory, BF theory, and Chern-Simons theory.}
  type-I \cite{VijayXcube} fracton models that do not have any gauge-invariant fractal operators \cite{Ifoot:fractal}.
The mobility constraints \cite{gauging} and other important properties \cite{ShirleyEntanglement,PaiFusion}
  of these models have a fundamental dependence on a layering structure of spacetime, known as a foliation structure \cite{3manifolds}, see \figref{fig:foliation}.
\refscite{defectNetworks,WenCellular,JuvenCellular,stringMembraneNet} have shown that these fracton phases can be thought of as a
  topological quantum field theory (TQFT) that is embedded with stacks of interfaces (also called defects) upon which certain anyons are condensed.
These interfaces are the so-called leaves (i.e. layers) of the foliation.
Therefore, instead of coupling to a metric $g_{\mu\nu}$, these fracton phases are coupled to one or more foliations.
For example, the X-cube model \cite{VijayXcube} on a simple cubic lattice is coupled to three flat foliations,
  but more generic foliations are also allowed \cite{3manifolds,SlagleLattices}.
This is in contrast to TQFT (without interfaces or defects),
  which does not couple to a metric or foliation.

Previous works have uncovered field theories for the X-cube and other gapped
  fracton models \cite{XcubeQFT,stringMembraneNet,Seiberg3,Seiberg4,ChamonCS,YizhiBF,YizhiChernSimon}.
In \refcite{stringMembraneNet}, the X-cube fracton model was generalized to manifolds with arbitrary curved foliations,
  but formally quantizing the field theory was left as an open problem.
\refcite{Seiberg3} later showed how to formally treat the X-cube field theory from \refcite{XcubeQFT} as a
  quantum field theory (QFT) with quantized coefficients.

In this work, we wish to quantize the foliated field theory from \refcite{stringMembraneNet}.
This task is nontrivial and requires new ideas,
  such as the introduction of a new kind of \emph{foliated} gauge field,
  which behaves like a stack of ordinary gauge fields.
We call a QFT with foliated gauge fields a \emph{foliated quantum field theory} (FQFT).

We also show that the FQFT is a \emph{foliated fracton order} \cite{Cfoot:foliatedOrder,3manifolds,twisted,ShirleyEntanglement}.
Foliated fracton orders have ground states for which a local unitary transformation can decouple 2D topological orders from the ground state.
In the FQFT, this transformation exhibits an IR duality that decouples 2+1D gauge theories from the FQFT
  by giving a coupling constant a piecewise spatial dependence
  which can be manipulated by the duality.

\arxiv{
In the following, we begin by reviewing how to mathematically describe a foliation using a 1-form foliation field.
We then introduce the FQFT and discuss its gauge invariant operators, level quantization, and foliated fracton order.
Some technical details and extended discussions appear in the \arxivPR{Appendix}{Supplementary Material (SM) \cite{supplementary}}.
See \refcite{talk} for a recorded talk.}

\mysec{Foliation Field}
\label{sec:foliation field}
A \emph{foliation} is a decomposition of a manifold into an infinite number of disjoint lower-dimensional submanifolds
  called \emph{leaves}.
A common example is to decompose 3+1D spacetime into 3D spatial slices;
  in this example, the codimension-1 leaves can be indexed by the time coordinate.

We will describe a codimension-1 foliation using a 1-form foliation field $e_\mu$.
The foliation field is analogous to a metric $g_{\mu\nu}$,
  except $e_\mu$ describes a foliation geometry instead of a Riemannian geometry.
The leaves of the foliation are defined to be the codimension-1 submanifolds that are orthogonal to the foliation field.
That is, the tangent vectors $v^\mu$ of the leaves are in the null space of the foliation field covector: $v^\mu e_\mu = 0$.
\arxivPR{In order for this definition to work, the}{The}
  foliation field must never be zero \arxiv{[$e(x) \neq 0 \;\, \forall x$] }and it must satisfy the following constraint\footnote{%
    We use differential form notation throughout this work.
    In components, \eqnref{eq:e constraint} can be written as
      $\epsilon^{\alpha\beta\gamma\delta} e_\beta \partial_\gamma e_\delta = 0$
      where $\epsilon$ is the Levi-Civita symbol.}:
\begin{equation}
  e \wedge d e = 0 \label{eq:e constraint}
\end{equation}
(which can be viewed as a special case of the Frobenius theorem \cite{FrankelGeometry}).

More intuition can be obtained by noting that the foliation is invariant under a ``gauge transformation'' that rescales the foliation field (since this does not affect orthogonality to the leaves):
\begin{equation}
  e \to \gamma e \label{eq:e transform}
\end{equation}
where $\gamma$ is a scalar function.
It is always possible to apply the above transformation such that within an open ball of spacetime,
  the foliation fields are closed $(d e = 0)$ and can be written as the derivative of a scalar function $f$: $e = df$.
Locally, $f$ can be thought of as a coordinate that indexes the leaves of the foliation,
  similar to how a time coordinate indexes time slices of spacetime.

To foliate a torus, the foliation field can be chosen to be closed (e.g. $e=dx$ so that $de=0$).
For more exotic foliations,
  the exterior derivative takes the form $d e = e \wedge \beta$
  \arxiv{[which satisfies \eqnref{eq:e constraint}] }%
  for some 1-form $\beta$.
The cohomology class of $\beta \wedge d\beta$ is the so-called Godbillon-Vey invariant of the foliation \cite{GodbillonVey,GodbillonVeyFamilies},
  which classifies the obstruction to a closed foliation field.
Under $e \to \gamma e$\arxiv{ [from \eqnref{eq:e transform}]},
  $\beta$ transforms as $\beta \to \beta - d \gamma$.

Multiple simultaneous foliations $e^k$ are indexed by the superscript $k = 1, 2, \ldots, \nf$ (\figref{fig:foliation}).
Each foliation satisfies \eqnref{eq:e constraint} independently: $e^k \wedge d e^k = 0$.
We never implicitly sum over repeated foliation indices $k$.

\mysec{Foliated QFT}
The foliated QFT (FQFT) Lagrangian is\footnote{%
  In components, $L = \Big[ \sum_{k=1}^{\nf} \frac{M_k}{2\pi} (\partial_\alpha B^k_\beta + n_k b_{\alpha\beta}) A^k_{\gamma\delta}
                      + \frac{N}{2\pi} b_{\alpha\beta} \partial_\gamma a_\delta \Big] \epsilon^{\alpha\beta\gamma\delta} \, \dd^4x$,
    and \eqnref{eq:A constraint} can be written as $\epsilon^{\alpha\beta\gamma\delta} A^k_{\beta\gamma} e^k_\delta = 0$.
  The Lagrangian can alternatively be written as
    $L = \sum_k \frac{M_k}{2\pi} B^k \wedge d A^k + \frac{N}{2\pi} b \wedge \left( d a + \sum_k m_k A^k \right)$.}
\begin{gather}
  L = \sum_{k = 1}^{\nf} \frac{M_k}{2\pi} (dB^k + n_k b) \wedge A^k + \frac{N}{2\pi} b \wedge da \label{eq:L} \\
  A^k \wedge e^k = 0 \label{eq:A constraint}
\end{gather}
$B^k$ and $a$ are 1-form gauge fields.
(Note that $B^k$ is not a magnetic field in this notation; $B^k$ has no dependence on $A^k$.)
$b$ is a 2-form gauge field.
$A^k$ are \emph{foliated} (1+1)-form gauge fields,
  which are locally 2-forms that obey the constraint \eqnref{eq:A constraint}.
\arxivPR{In \secref{sec:quantize}, we}{We} will show that
  the physics is equivalent under $n_k \sim n_k + N$ and that
  $M_k, n_k, N \in \ZZ$ are quantized level coefficients
  with $m_k \equiv \frac{n_k M_k}{N} \in \ZZ$
(and $M_k \neq 0$ and $N \neq 0$).
$\sum_{k = 1}^{\nf}$ sums over the different foliations.
Unlike the dynamical gauge fields ($A^k$, $B^k$, $a$, $b$),
  the foliation field $e_\mu$ is non-dynamical and is not integrated over in the partition function
  (analogous to a static metric $g_{\mu\nu}$).
Similar to a TQFT, FQFT does not couple to a metric.

If $n_k=0$, the second term in $L$ describes a 3+1D BF theory
  (which is a field theory for $Z_N$ gauge theory or 3D toric code\arxivPR{ \cite{Dfoot:BF}}{\cite{XcubeQFT,SeibergBFZn}}),
  while the first term is an FQFT for a stack of infinitesimally-spaced 2+1D BF theories for each foliation,
  (i.e. a field theory for stacks of $Z_{M_k}$ toric codes \cite{KitaevAnyon}).
When $M_k = N$ and $n_k=1$, the leaves are coupled to the 3+1D BF theory,
  and the resulting theory describes the ground state Hilbert space\footnote{%
    There are no excitations in the FQFT;
      the FQFT Hilbert space only consists of degenerate ground states.
    The same is true of BF theory (or Chern-Simons theory),
      which describes the ground state Hilbert space of toric code.
    However, string operators \arxiv{(which we study in \secref{sec:operators})} can be thought of as moving particles around spacetime.}
  of the $Z_N$ X-cube model \cite{XcubeQFT,Seiberg3,VijayXcube,stringMembraneNet} on any foliation \cite{Jfoot:XcubeFoliation}
  in the limit of infinitesimal lattice spacing.
This equivalence can be demonstrated in a number of ways \cite{Gfoot:XCubeEquiv} and will be exemplified in \arxivPR{\secref{sec:operators}}{this work}.
Some intuition from coupled-layer constructions of fracton models applies here as well
  \cite{VijayLayer,MaLayer,CageNet}.
A lattice model for this FQFT was given in Sec. 3 and Appendix A of \refcite{stringMembraneNet}.

\mysubsec{Foliated Gauge Field}
The foliated QFT includes a new kind a gauge field: a \emph{foliated} (1+1)-form gauge field $A^k$ for each foliation $k$.
A foliated (1+1)-form gauge field\arxiv{\footnote{%
  More generally, one could consider a foliated ($p+q$)-form gauge field that
    behaves similarly to independent $p$-form gauge fields on a codimension-$q$ foliation.}}
  behaves similarly to a stack of independent 1-form gauge fields.
This is desirable because when $n_k=0$, the first term in \eqnref{eq:L} should describe a stack of independent 2+1D gauge theories.

Locally, a foliated (1+1)-form gauge field $A^k$ is a 2-form gauge field that obeys the constraint \eqnref{eq:A constraint}.
Similar to ordinary gauge fields, the exterior derivative $dA^k$ is required to be well-defined.
Note that this requirement does not put any restriction on the continuity of the foliated gauge field $A^k$ between leaves of the foliation.
For example if $e^1=dz$, then the constraint \eqnref{eq:A constraint} implies that
  $A^1 = \tilde{A}^1 \wedge dz$ for some 1-form $\tilde{A}^1$,
  and $\tilde{A}^1$ can have arbitrary discontinuities in the z-direction
  since these discontinuities will not contribute to $dA^1$
  (due to the antisymmetry induced by the wedge product).
Furthermore, we allow foliated gauge fields to contain a delta-function onto a leaf.
For example if $e^1=dz$, then $A^1 = x \, \delta(z) \, dy \wedge dz$ is allowed.
See \apprefFst{app:foliated gauge field}{B} for a more formal definition of foliated gauge fields.

Since the first term in \eqnref{eq:L} should describe a stack of 2+1D BF theories for each $k$ with  $n_k=0$,
  the gauge fields $A^k$ and $B^k$ should effectively have three components
  (since the 1-form gauge fields in 2+1D BF theory have three components).
Considering again the example $e^1=dz$,
  we indeed see that the constraint \eqnref{eq:A constraint} implies that the foliated (1+1)-form has exactly three components:
  $A^1 = (A_{03}^1 \,\dd t + A_{13}^1 \,\dd x + A_{23}^1 \,\dd y) \wedge \dd z$.
The 1-form gauge field $B^k$ has four components ($B^1 = B_0^1 dt + B_1^1 dx + B_2^1 dy + B_3^1 dz$).
However there is a gauge symmetry $B^k \to \alpha^k$ for an arbitrary foliated (0+1)-form $\alpha^k$,
  which locally satisfies $\alpha^k \wedge e^k = 0$
  (i.e. locally $\alpha^k = \tilde{\alpha}^k e^k$ for some scalar $\tilde{\alpha}^k$).
This makes the $dz$ component an unimportant gauge redundancy.
Therefore, $A^k$ and $B^k$ both effectively have three components (for each foliation $k$), as desired.

\mysubsec{Fractons and Gauge Invariant Operators}
\label{sec:operators}
The set of gauge symmetries determines the set of gauge invariant operators.
In ordinary topological QFT (e.g. Chern-Simons theory),
  gauge invariant operators can be smoothly deformed into any shape.
However, in a foliated QFT, the gauge invariant operators are often constrained to the intersection of one or more leaves of different foliations.

Gauge invariant operators can be interpreted as moving topological excitations around in spacetime.
Therefore, the rigidity of the gauge invariant operators is analogous to the mobility constraints of the fracton, lineon, and planon particles.

The gauge transformations of the FQFT are
\begin{align}
\begin{split}
  A^k &\to A^k + d\zeta^k \\
  B^k &\to B^k + d\chi^k - n_k\mu + \alpha^k \\
  a   &\to a   + d\lambda - \sum_k m_k \zeta^k \\
  b   &\to b   + d\mu \label{eq:gauge}
\end{split}
\end{align}
where $m_k \equiv \frac{n_k M_k}{N}$.
$\chi^k$ and $\lambda$ are arbitrary 0-form gauge fields,
  while $\mu$ is an arbitrary 1-form gauge field.
$\zeta^k$ and $\alpha^k$ are foliated (0+1)-form gauge fields.
Locally, $\zeta^k$ are 1-form gauge fields that satisfy the constraint\footnote{%
  A more relaxed constraint $d \zeta^k \wedge e^k = 0$ is sufficient to guarantee that $A^k \wedge e^k = 0$ is preserved.
  However, this relaxed constraint would allow transformations such as $\zeta^k = c \, dz$ for any $c \in \RR$ (i.e. $c$ is not quantized)
    on a manifold with periodic boundaries in $z$.
  This would be undesirable since it would make \eqnref{eq:fracton} not gauge invariant,
    which would be inconsistent with the lattice model version of this theory \cite{stringMembraneNet}.
  }
  $\zeta^k \wedge e^k = 0$, and similar for $\alpha^k$.

Consider the following string operator:
\begin{equation}
  W = e^{\ii\, q \oint_{\M_1^\text{F}} a} \label{eq:fracton}
\end{equation}
where $\M_1^\text{F}$ is a 1-dimensional manifold described below.
Large gauge transformations imply that the charge $q$ is an integer.
A nonlocal ``equation of motion'' (from integrating out $b$) shows that $W=1$ when $q$ is an integer multiple of $N$ \cite{Ffoot:appEoM}.
Therefore $W$ only depends on $q$ modulo $N$.
After a gauge transformation, $W \to W \exp\!\left[\ii \oint_{\M_1^\text{F}} \left( d\lambda - \sum_k m_k \zeta^k\right)\right]$.
The first term, $\oint_{\M_1^\text{F}} d\lambda$, is invariant if $\M_1^\text{F}$ is a closed loop.
The second term, $\oint_{\M_1^\text{F}} \sum_k m_k \zeta^k$,
  is invariant if the tangent vectors $v^\mu$ of $\M_1^\text{F}$ are in the null space of each $m_k \zeta^k$,
  i.e. $v^\mu m_k \zeta^k_\mu = 0$.
But locally, $\zeta^k_\mu = \tilde{\zeta}^k e^k_\mu$ for some scalar $\tilde{\zeta}^k$.
Therefore the second term is gauge invariant if for each $k$ with $m_k \neq 0$,
  the loop $\M_1^\text{F}$ is supported on a single leaf of the $k^\mathrm{th}$ foliation
  [since then $v^\mu m_k \zeta^k_\mu \propto v^\mu e^k_\mu$ and $v^\mu e^k_\mu = 0$ by the definition of $e^k$ above \eqnref{eq:e constraint}].

Therefore, if there are $n$ foliations with $m_k \neq 0$,
  then the string operator [\eqnref{eq:fracton}] and the particle it transports
  are bound to the intersection of $n$ leaves.
If there are three or more spatial\footnote{%
  A spatial foliation is a foliation that has no time component,
    i.e. $e^k(\hat{t}) = e^k_\mu \hat{t}^\mu = 0$
    where $\hat{t}^\mu$ is a vector pointing in the time direction.}
  foliations (that are spatially transverse\footnote{%
  By spatially transverse, we mean that the spatial foliation of 2D leaves is transverse.
  Transverse means that when $n$ leaves intersect at a point,
      then the intersection of the tangent spaces of the $n$ leaves at this point is just the null vector. \cite{totalFoliation}})
  as in \figref{fig:foliation},
  then this string operator can move fractons in time (assuming time is periodic),
  but it cannot move fractons spatially.
When $n_k = 1$ and $M_k = N$, this fracton is equivalent to the X-cube fracton \cite{VijayXcube} for any foliation\cite{Jfoot:XcubeFoliation}.
It has been proven that all compact orientable 3-manifolds admit a total foliation (i.e. three transverse foliations) \cite{totalFoliation},
  which implies that all such manifolds admit an FQFT with fractons.

Consider a different string operator:
\begin{equation}
  T = e^{\ii \oint_{\M_1^\text{L}} \sum_k q_k B^k} \label{eq:lineon}
\end{equation}
Large gauge transformations imply that the charges $q_k \in \ZZ$ are integers.
The $B^k \to B^k + d\chi^k$ gauge transformation shows that $\M_1^\text{L}$ must be be a closed loop.
The $B^k \to B^k + \alpha^k$ gauge transformation
  [where $\alpha^k = \tilde{\alpha}^k e^k$ (locally) is a foliated (0+1)-form]
  shows that $\M_1^\text{L}$ is supported on the intersection of $n$ leaves,
  where $n$ is the number of foliations $k$ with nonzero $q_k \neq 0$.
Finally, the $B^k \to B^k - n_k \mu$ gauge transformation implies that $\sum_k q_k n_k = 0$.
Therefore, the set of allowed charge vectors forms an abelian group $G = \{q \in \ZZ^{\nf} \;|\; \sum_k q_k n_k = 0\}$.

\begin{figure}
  \subfloat[\label{fig:fracton}]{\includegraphics[width=.33\columnwidth]{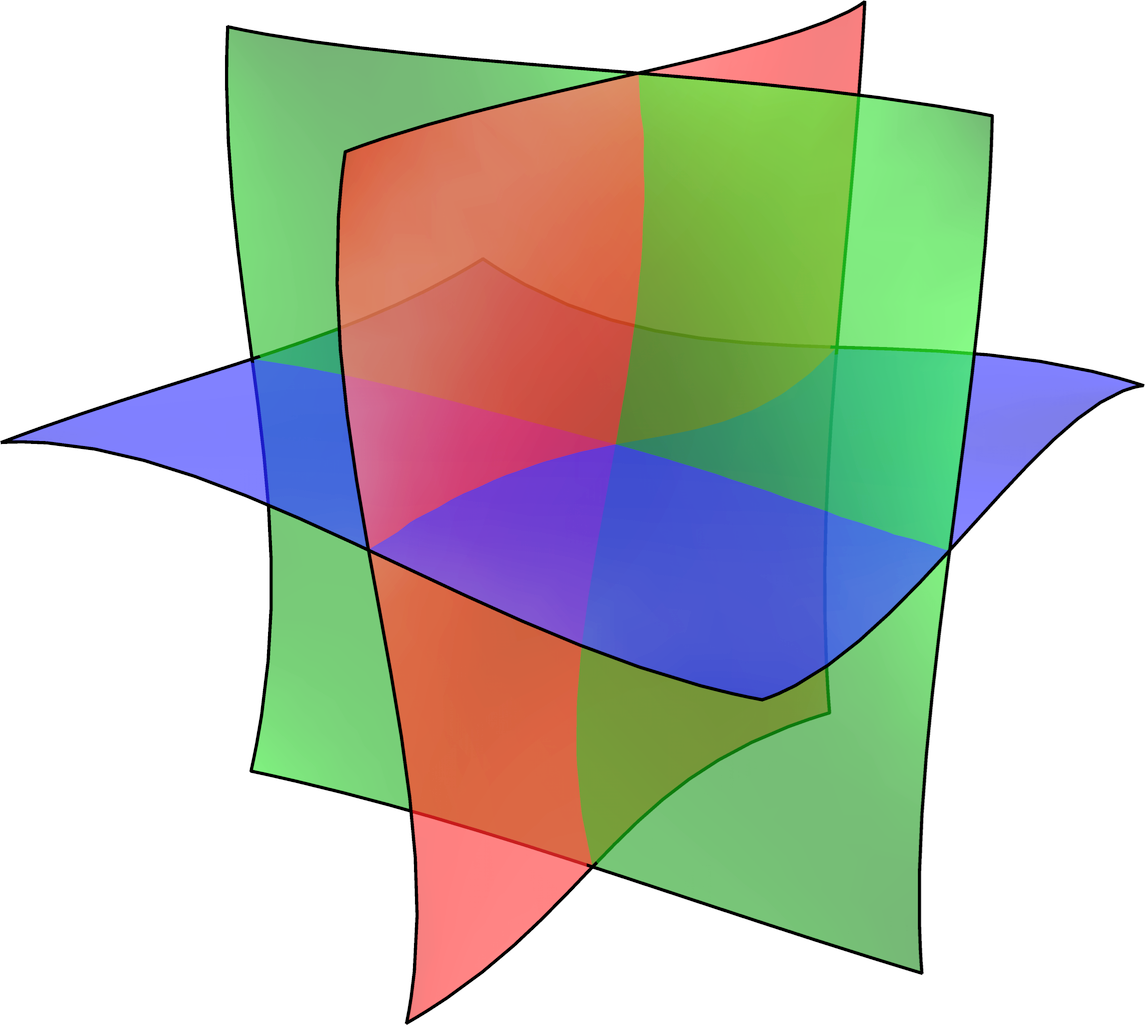}}  \hspace{.02\columnwidth}
  \subfloat[\label{fig:lineon}]{\includegraphics[width=.22\columnwidth]{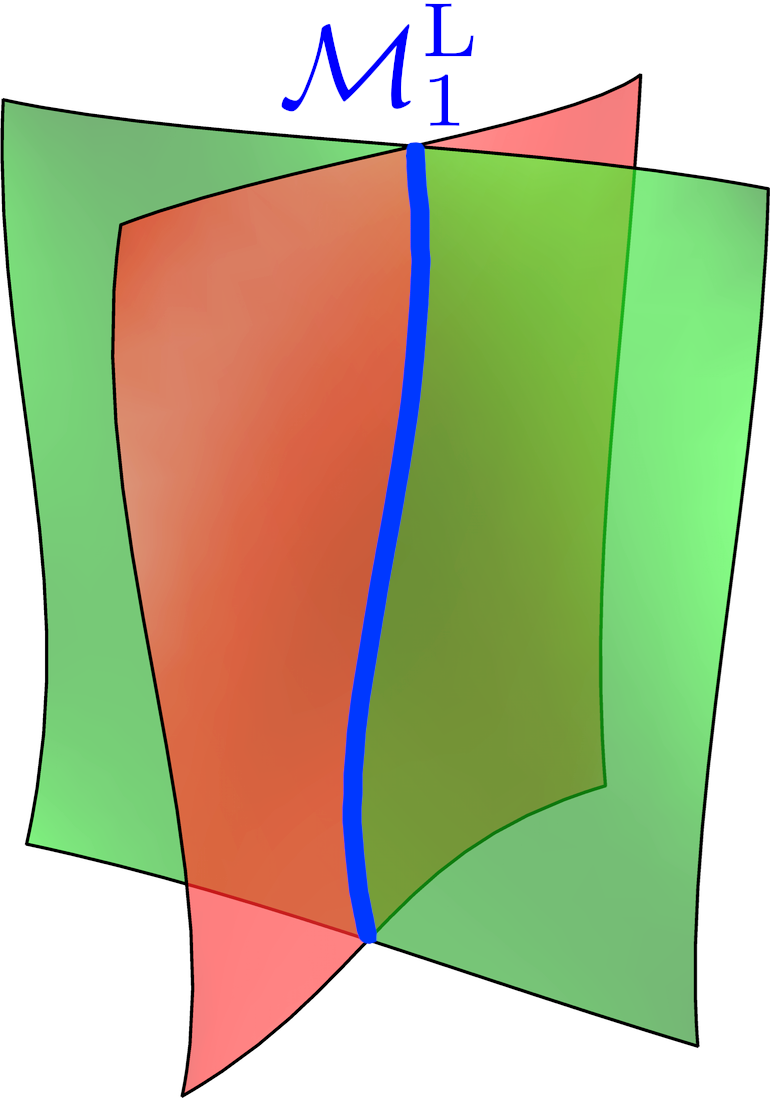}} \hspace{.02\columnwidth}
  \subfloat[\label{fig:planon}]{\raisebox{.5cm}{\includegraphics[width=.38\columnwidth]{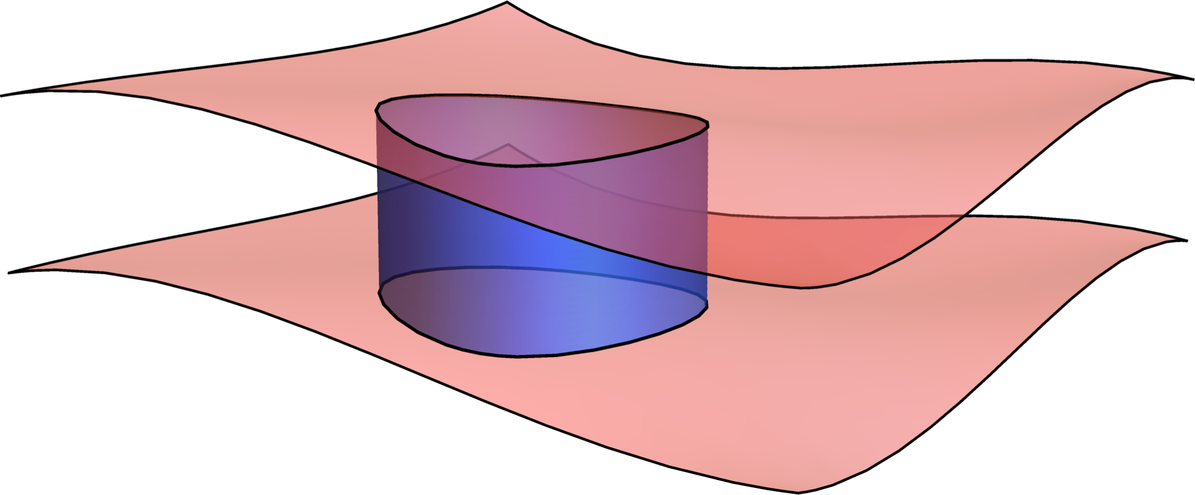}}}
  \caption{%
    Spatial pictures of:
    {\bf (a)} Three leaves intersecting at a point.
    {\bf (b)} A 1-dimensional manifold $\M_1^\text{L}$ (blue) at the intersection of two leaves (red and green).
    {\bf (c)} A 2-dimensional manifold $\M_2^\text{P}$ (blue) with boundaries supported on leaves (red).
  }
\end{figure}

A nonlocal ``equation of motion'' (from integrating out $A^k$) shows \cite{Ffoot:appEoM} that $T=1$ when $q_k \in M_k \ZZ$ (and $\sum_k q_k n_k = 0$).
Thus, the trivial charge vectors form a subgroup $N = \{q \in G \;|\; q_k \in M_k \ZZ \} \triangleleft G$.
Since both of these groups ($G$ and $N$) are isomorphic to $\ZZ^{\nf-1}$ (or $\ZZ^{\nf}$ if $n_k=0$),
  their quotient group $G/N$ of physically distinct charge vectors is a finite abelian group
  (i.e. $G/N$ is isomorphic to $Z_{r_1} \times \cdots \times Z_{r_{\nf-1}}$ for some integers $r_i \in \NN$).

In the $Z_N$ X-cube model example with three foliations and $n_k = 1$ and $M_k = N$,
  the allowed charge vectors are spanned by $q_k^\text{(X)} = (0,1,-1)$ and $q_k^\text{(Z)} = (1,-1,0)$.
These particles are bound to a pair of leaves (\figref{fig:lineon})
  and are therefore restricted to spatially only move along 1D lines (for spatial foliations).
These are $Z_N$ X-cube lineons.
For the standard three flat foliations ($e^1=\dd x$, $e^2=\dd y$, $e^3=\dd z$),
  $q_k^\text{(X)}$ and $q_k^\text{(Z)}$ can move only in the X and Z directions, respectively;
  and their sum $q_k^\text{(X)} + q_k^\text{(Z)} = (1,0,-1)$ can only move in the Z direction.
This is analogous to the X-cube model where the composition of an X-axis lineon with a Z-axis lineon is a Y-axis lineon.
The physics generalizes naturally to $n$ foliations:
  the charge vectors are spanned by $n-1$ vectors of the form $(\cdots,0,1,-1,0,\cdots)$.
Note that even if a charge vector has three nonzero components,
  it is not a fracton;
  instead, it is the composition of at most two lineons.

Even for an arbitrary number of foliations and coefficients $n_k$,
  it is always possible to decompose a charge vector $q_k$ into lineon and planon charges
  (which have at most two nonzero elements $q_k \neq 0$).
See \appref{app:lineon}{E} for a proof.
Therefore, the string operator $T$ only describes lineons (or composites of lineons and planons),
  but never fractons.

Other gauge invariant operators include
\begin{align}
  T' = e^{\ii \oint_{\M_2} b} &&
  W' = e^{\ii \oint_{\M_2^\text{P}} A^k} \label{eq:T'W'}
\end{align}
$\oint_{\M_2} b$ denotes an integral of $b$ over a closed 2-manifold $\M_2$.
$\oint_{\M_2^\text{P}} A^k$ denotes an integral of $A^k$ over a 2-manifold $\M_2^\text{P}$
  with boundaries that must each be supported on a single leaf of the foliation $k$, as in \figref{fig:planon}.
$T'$ wraps a string excitation around $\M_2$.
In the X-cube model example, $T'$ measures the number of fractons inside $\M_2$.
In the X-cube lattice model, this operator is a complicated operator that
  wraps a loop of many lineon excitations around $\M_2$.\footnote{%
    Although an isolated lineon can only move along a straight line without creating additional excitations,
      a string of many lineons can move more freely, especially when additional excitations are allowed to be created.}
In the X-cube example, $W'$ moves a pair of X-cube fractons\footnote{%
  A pair of X-cube fractons can form a planon, which has 2D mobility.}
  around the top and bottom boundaries of the blue 2-manifold $\M_2^\text{P}$ shown in \figref{fig:planon}.

See \appref{app:mobility}{C} for more general operators and a different approach to understanding the particle mobility constraints.

\mysubsec{Level Quantization}
\label{sec:quantize}
Now we study the quantization of the level coefficients $M_k$, $n_k$, and $N$.
First note that $m_k \equiv \frac{n_k M_K}{N}$ and $n_k$ appear as coefficients in
  the gauge transformations [\eqnref{eq:gauge}] of compact gauge fields ($a$ and $B^k$).
This implies that $m_k, n_k \in \ZZ$.

The Lagrangian transforms under the gauge transformations\arxiv{ [\eqnref{eq:gauge}]} as:
\begin{equation}
\begin{aligned}
  L \to L' = L &+ \sum_k \frac{M_k}{2\pi} (d B^k \wedge d\zeta^k + d\chi^k \wedge d A^k) \\
               &+ \frac{N}{2\pi} (d b \wedge d\lambda + d\mu \wedge d a)
\end{aligned}
\end{equation}
Locally, the new terms are total derivatives.
But since these are derivatives of gauge fields,
  their integral over a closed manifold can be nonzero.
However, the integral is quantized such that the change in the action is
  an integer multiple of $2\pi M_k$ plus an integer multiple of $2\pi N$.
Therefore, the partition function $Z = e^{i \int L}$ is gauge invariant if $M_k, N \in \ZZ$.

The equations of motion that result from integrating out $a$ and $B^k$
  imply \cite{Ffoot:appEoM} that locally $db = dA^k = 0$
  and globally the operators in \eqnref{eq:T'W'} are quantized:
\begin{align}
  \oint_{\M_2} b &\in \frac{2\pi}{N}\ZZ &
  \oint_{\M_2^\text{P}} A^k &\in \frac{2\pi}{M_k}\ZZ \label{eq:quantized EoM}
\end{align}
Together, these local and global equations of motion show that the $b \wedge A^k$ term
  in the FQFT action [\eqnref{eq:L}] is quantized as follows:
\begin{equation}
  \frac{M_k n_k}{2\pi} \int b \wedge A^k \in 2\pi \frac{n_k}{N} \ZZ
\end{equation}
This implies that the action is invariant under the following identification $n_k \sim n_k + N$.

\mysec{Exfoliation}
\label{sec:exfoliation}
\refcite{3manifolds} showed that a finite-depth local unitary transformation can map between the ground states of
  (1) an X-cube model of lattice length $L_0$ in one direction, and
  (2) an X-cube model of lattice length $L_0-1$ in the same direction
      along with a decoupled layer of toric code
      (and some trivial decoupled qubits).
We will refer to this process as \emph{exfoliation}.
In high-energy terminology, exfoliation corresponds to an IR duality
  that decouples 2+1D gauge theories from a 3D FQFT.
A fracton order that admits exfoliation is said to be a \emph{foliated fracton order} \cite{Cfoot:foliatedOrder,3manifolds,twisted}.
The X-cube model is a foliated fracton order that is foliated by toric code layers \cite{3manifolds,Afoot:twisted}.

We now show that the FQFT is a foliated fracton order by exfoliating 2+1D BF theories.
For simplicity, consider a flat foliation $e^1 = dz$
  (which may coexist with other foliations $e^k$).
We want to demonstrate a duality from an FQFT with constant $n_1 \in \ZZ$
  to an FQFT with a spatially-dependent $\tilde{n}_1(z)$ that is zero within $z_1 < z < z_2$:
\begin{equation}
  n_1 \leftrightarrow \tilde{n}_1(z) = \begin{cases} n_1 & z \leq z_1 \text{ or } z \geq z_2 \\
                                         0   & z_1 < z < z_2 \end{cases} \label{eq:n duality}
\end{equation}
On the right-hand-side of the duality,
  the $A^1$ and $B^1$ fields within $z_1 < z < z_2$ are decoupled from the rest of the fields.
The equations of motion for $A^1$ and $B^1$ are $dA^1 = dB^1 \wedge e^1 = 0$ within $z_1 < z < z_2$.
These equations of motion do not contain $z$-derivatives $\partial_z$ [recall $A^1 \wedge e^1 = 0$ from \eqnref{eq:A constraint}],
  which shows that $A^1$ and $B^1$ at different $z$ are completely decoupled.
These decoupled fields constitute an exfoliated stack of infinitesimally-spaced 2+1D BF theories.

The duality results from the following transformation:
\begin{align}
  a   &\leftrightarrow \tilde{a}   =
    \begin{cases} a & z \leq z_1 \text{ or } z_2 \leq z \\
                  a + m_1 \int_{z_1}^z A^1 & z_1 < z < z_2 \end{cases} \nn\\
  A^1 &\leftrightarrow \tilde{A}^1 = A^1 + \delta(z-z_2) \int_{z_1}^{z_2} \dd z \, A^1 \label{eq:aB duality}\\
  B^1 &\leftrightarrow \tilde{B}^1 =
    \begin{cases} B^1 & z \leq z_1 \text{ or } z_2 \leq z \\
                  B^1 - B^1(z_2) + n_1 \int_z^{z_2} b & z_1 < z < z_2 \end{cases} \nn
\end{align}
We are using a notation where the integrals above are defined as
  $\big( \int_{z_1}^z A^1 \big)_\mu \equiv \int_{z_1}^z A^1_{3\mu} \, \dd z$,
  $\big( \int_{z_1}^{z_2} \dd z \, A^1 \big)_{\mu\nu} = \int_{z_1}^{z_2} \dd z \, A^1_{\mu\nu}$, and
  $\left( \int_z^{z_2} b \right)_\mu \equiv \int_z^{z_2} b_{3\mu} \, \dd z$.\footnote{%
  In \eqnref{eq:aB duality}, we use the convention that integrals do not pick up delta functions on their end points;
    e.g. $\int_0^1 \delta(x) \dd x = 0$.}
In order for this definition to make sense,
  we have implicitly chosen a flat connection to parallel transport the gauge fields.
$B^1(z_2)$ is shorthand for $B^1(t,x,y,z_2)$,
  just as $B^1$ is shorthand for $B^1(t,x,y,z)$.
In \appref{app:exfoliation}{G.1}, we show that the above equation transforms the
  equations of motion according to \eqnref{eq:n duality},
  which demonstrates the exfoliation duality.

Note that since the duality acts nonlocally on the fields,
  the locality of some gauge invariant operators can change.
Indeed, this must occur because $\tilde{n}_1 = 0$ will result in less rigidity constraints on the gauge invariant operators.
See \appref{app:exfoliation}{G} for examples and further discussion.

\mysec{Conclusion}
We have introduced a generic foliated QFT (FQFT) that is capable of describing a large class of foliated gapped fracton models on foliated manifolds.
We also demonstrated a novel duality that spatially transforms the level coefficients,
  which shows that the FQFT is a foliated fracton order \cite{Cfoot:foliatedOrder,3manifolds,twisted}.

Many future directions remain.
Additional terms can be added to the FQFT Lagrangian
  to realize more exotic fracton models
  \cite{twisted,WilliamsonNonAbelianDefects,PremPermutationNonAbelian,BulmashNonAbelian,DuaSSET,TantivasadakarnSymmetryEnriched,ShirleySSPTdual}.
The fractonic Higgs mechanism \cite{BulmashHiggs,MaHiggs} could be revisited now that we understand
  gapped fracton orders on curved foliations
  \cite{3manifolds,stringMembraneNet,defectNetworks,SystematicRadicevic}
  and $U(1)$ fracton models
  \cite{PretkoU1,RasmussenU1,PretkoEM,RadzihovskyVector,SeibergU1,SeibergSymmetry,PretkoTemperature}
  on curved space \cite{SlagleCurvedU1}.
Finally, FQFT could provide further insight on other works,
  such as the study of boundaries of fracton models \cite{BulmashBoundary}
  or models in higher dimensions \cite{PengYeDimensions}.

\begin{acknowledgments}
We thank Shu-Heng Shao, Nathan Seiberg, Ho Tat Lam, Pranay Gorantla, Po-Shen Hsin, Anton Kapustin, Xie Chen, and Wilbur Shirley for helpful discussion.
K.S. is supported by the Walter Burke Institute for Theoretical Physics at Caltech.
\end{acknowledgments}

\bibliography{FQFT}

\begin{thebibliography}{102}%
\makeatletter
\providecommand \@ifxundefined [1]{%
 \@ifx{#1\undefined}
}%
\providecommand \@ifnum [1]{%
 \ifnum #1\expandafter \@firstoftwo
 \else \expandafter \@secondoftwo
 \fi
}%
\providecommand \@ifx [1]{%
 \ifx #1\expandafter \@firstoftwo
 \else \expandafter \@secondoftwo
 \fi
}%
\providecommand \natexlab [1]{#1}%
\providecommand \enquote  [1]{``#1''}%
\providecommand \bibnamefont  [1]{#1}%
\providecommand \bibfnamefont [1]{#1}%
\providecommand \citenamefont [1]{#1}%
\providecommand \href@noop [0]{\@secondoftwo}%
\providecommand \href [0]{\begingroup \@sanitize@url \@href}%
\providecommand \@href[1]{\@@startlink{#1}\@@href}%
\providecommand \@@href[1]{\endgroup#1\@@endlink}%
\providecommand \@sanitize@url [0]{\catcode `\\12\catcode `\$12\catcode
  `\&12\catcode `\#12\catcode `\^12\catcode `\_12\catcode `\%12\relax}%
\providecommand \@@startlink[1]{}%
\providecommand \@@endlink[0]{}%
\providecommand \url  [0]{\begingroup\@sanitize@url \@url }%
\providecommand \@url [1]{\endgroup\@href {#1}{\urlprefix }}%
\providecommand \urlprefix  [0]{URL }%
\providecommand \Eprint [0]{\href }%
\providecommand \doibase [0]{http://dx.doi.org/}%
\providecommand \selectlanguage [0]{\@gobble}%
\providecommand \bibinfo  [0]{\@secondoftwo}%
\providecommand \bibfield  [0]{\@secondoftwo}%
\providecommand \translation [1]{[#1]}%
\providecommand \BibitemOpen [0]{}%
\providecommand \bibitemStop [0]{}%
\providecommand \bibitemNoStop [0]{.\EOS\space}%
\providecommand \EOS [0]{\spacefactor3000\relax}%
\providecommand \BibitemShut  [1]{\csname bibitem#1\endcsname}%
\let\auto@bib@innerbib\@empty
\bibitem [{\citenamefont {{Nandkishore}}\ and\ \citenamefont
  {{Hermele}}(2019)}]{FractonRev1}%
  \BibitemOpen
  \bibfield  {author} {\bibinfo {author} {\bibfnamefont {R.~M.}\ \bibnamefont
  {{Nandkishore}}}\ and\ \bibinfo {author} {\bibfnamefont {M.}~\bibnamefont
  {{Hermele}}},\ }\href {\doibase 10.1146/annurev-conmatphys-031218-013604}
  {\bibfield  {journal} {\bibinfo  {journal} {Annual Review of Condensed Matter
  Physics}\ }\textbf {\bibinfo {volume} {10}},\ \bibinfo {pages} {295}
  (\bibinfo {year} {2019})},\ \Eprint {http://arxiv.org/abs/1803.11196}
  {arXiv:1803.11196} \BibitemShut {NoStop}%
\bibitem [{\citenamefont {{Pretko}}\ \emph {et~al.}(2020)\citenamefont
  {{Pretko}}, \citenamefont {{Chen}},\ and\ \citenamefont
  {{You}}}]{FractonRev2}%
  \BibitemOpen
  \bibfield  {author} {\bibinfo {author} {\bibfnamefont {M.}~\bibnamefont
  {{Pretko}}}, \bibinfo {author} {\bibfnamefont {X.}~\bibnamefont {{Chen}}}, \
  and\ \bibinfo {author} {\bibfnamefont {Y.}~\bibnamefont {{You}}},\ }\href
  {\doibase 10.1142/S0217751X20300033} {\bibfield  {journal} {\bibinfo
  {journal} {International Journal of Modern Physics A}\ }\textbf {\bibinfo
  {volume} {35}},\ \bibinfo {eid} {2030003} (\bibinfo {year} {2020})},\ \Eprint
  {http://arxiv.org/abs/2001.01722} {arXiv:2001.01722} \BibitemShut {NoStop}%
\bibitem [{\citenamefont {{Vijay}}\ \emph {et~al.}(2015)\citenamefont
  {{Vijay}}, \citenamefont {{Haah}},\ and\ \citenamefont
  {{Fu}}}]{VijayFracton}%
  \BibitemOpen
  \bibfield  {author} {\bibinfo {author} {\bibfnamefont {S.}~\bibnamefont
  {{Vijay}}}, \bibinfo {author} {\bibfnamefont {J.}~\bibnamefont {{Haah}}}, \
  and\ \bibinfo {author} {\bibfnamefont {L.}~\bibnamefont {{Fu}}},\ }\href
  {\doibase 10.1103/PhysRevB.92.235136} {\bibfield  {journal} {\bibinfo
  {journal} {\prb}\ }\textbf {\bibinfo {volume} {92}},\ \bibinfo {eid} {235136}
  (\bibinfo {year} {2015})},\ \Eprint {http://arxiv.org/abs/1505.02576}
  {arXiv:1505.02576} \BibitemShut {NoStop}%
\bibitem [{\citenamefont {{Vijay}}\ \emph {et~al.}(2016)\citenamefont
  {{Vijay}}, \citenamefont {{Haah}},\ and\ \citenamefont {{Fu}}}]{VijayXcube}%
  \BibitemOpen
  \bibfield  {author} {\bibinfo {author} {\bibfnamefont {S.}~\bibnamefont
  {{Vijay}}}, \bibinfo {author} {\bibfnamefont {J.}~\bibnamefont {{Haah}}}, \
  and\ \bibinfo {author} {\bibfnamefont {L.}~\bibnamefont {{Fu}}},\ }\href
  {\doibase 10.1103/PhysRevB.94.235157} {\bibfield  {journal} {\bibinfo
  {journal} {\prb}\ }\textbf {\bibinfo {volume} {94}},\ \bibinfo {eid} {235157}
  (\bibinfo {year} {2016})},\ \Eprint {http://arxiv.org/abs/1603.04442}
  {arXiv:1603.04442} \BibitemShut {NoStop}%
\bibitem [{\citenamefont {{Haah}}(2011)}]{HaahCode}%
  \BibitemOpen
  \bibfield  {author} {\bibinfo {author} {\bibfnamefont {J.}~\bibnamefont
  {{Haah}}},\ }\href {\doibase 10.1103/PhysRevA.83.042330} {\bibfield
  {journal} {\bibinfo  {journal} {\pra}\ }\textbf {\bibinfo {volume} {83}},\
  \bibinfo {eid} {042330} (\bibinfo {year} {2011})},\ \Eprint
  {http://arxiv.org/abs/1101.1962} {arXiv:1101.1962} \BibitemShut {NoStop}%
\bibitem [{\citenamefont {{Bravyi}}\ \emph {et~al.}(2011)\citenamefont
  {{Bravyi}}, \citenamefont {{Leemhuis}},\ and\ \citenamefont
  {{Terhal}}}]{BravyiFracton}%
  \BibitemOpen
  \bibfield  {author} {\bibinfo {author} {\bibfnamefont {S.}~\bibnamefont
  {{Bravyi}}}, \bibinfo {author} {\bibfnamefont {B.}~\bibnamefont
  {{Leemhuis}}}, \ and\ \bibinfo {author} {\bibfnamefont {B.~M.}\ \bibnamefont
  {{Terhal}}},\ }\href {\doibase 10.1016/j.aop.2010.11.002} {\bibfield
  {journal} {\bibinfo  {journal} {Annals of Physics}\ }\textbf {\bibinfo
  {volume} {326}},\ \bibinfo {pages} {839} (\bibinfo {year} {2011})},\ \Eprint
  {http://arxiv.org/abs/1006.4871} {arXiv:1006.4871} \BibitemShut {NoStop}%
\bibitem [{\citenamefont {Bravyi}\ and\ \citenamefont
  {Haah}(2013)}]{HaahMemory}%
  \BibitemOpen
  \bibfield  {author} {\bibinfo {author} {\bibfnamefont {S.}~\bibnamefont
  {Bravyi}}\ and\ \bibinfo {author} {\bibfnamefont {J.}~\bibnamefont {Haah}},\
  }\href {\doibase 10.1103/PhysRevLett.111.200501} {\bibfield  {journal}
  {\bibinfo  {journal} {Phys. Rev. Lett.}\ }\textbf {\bibinfo {volume} {111}},\
  \bibinfo {pages} {200501} (\bibinfo {year} {2013})},\ \Eprint
  {http://arxiv.org/abs/1112.3252} {arXiv:1112.3252} \BibitemShut {NoStop}%
\bibitem [{\citenamefont {{Brown}}\ \emph {et~al.}(2016)\citenamefont
  {{Brown}}, \citenamefont {{Loss}}, \citenamefont {{Pachos}}, \citenamefont
  {{Self}},\ and\ \citenamefont {{Wootton}}}]{BrownQuantumMemory}%
  \BibitemOpen
  \bibfield  {author} {\bibinfo {author} {\bibfnamefont {B.~J.}\ \bibnamefont
  {{Brown}}}, \bibinfo {author} {\bibfnamefont {D.}~\bibnamefont {{Loss}}},
  \bibinfo {author} {\bibfnamefont {J.~K.}\ \bibnamefont {{Pachos}}}, \bibinfo
  {author} {\bibfnamefont {C.~N.}\ \bibnamefont {{Self}}}, \ and\ \bibinfo
  {author} {\bibfnamefont {J.~R.}\ \bibnamefont {{Wootton}}},\ }\href {\doibase
  10.1103/RevModPhys.88.045005} {\bibfield  {journal} {\bibinfo  {journal}
  {Reviews of Modern Physics}\ }\textbf {\bibinfo {volume} {88}},\ \bibinfo
  {eid} {045005} (\bibinfo {year} {2016})},\ \Eprint
  {http://arxiv.org/abs/1411.6643} {arXiv:1411.6643} \BibitemShut {NoStop}%
\bibitem [{\citenamefont {{Chamon}}(2005)}]{ChamonGlassy}%
  \BibitemOpen
  \bibfield  {author} {\bibinfo {author} {\bibfnamefont {C.}~\bibnamefont
  {{Chamon}}},\ }\href {\doibase 10.1103/PhysRevLett.94.040402} {\bibfield
  {journal} {\bibinfo  {journal} {\prl}\ }\textbf {\bibinfo {volume} {94}},\
  \bibinfo {eid} {040402} (\bibinfo {year} {2005})},\ \Eprint
  {http://arxiv.org/abs/cond-mat/0404182} {arXiv:cond-mat/0404182} \BibitemShut
  {NoStop}%
\bibitem [{\citenamefont {{Prem}}\ \emph {et~al.}(2017)\citenamefont {{Prem}},
  \citenamefont {{Haah}},\ and\ \citenamefont {{Nandkishore}}}]{PremGlassy}%
  \BibitemOpen
  \bibfield  {author} {\bibinfo {author} {\bibfnamefont {A.}~\bibnamefont
  {{Prem}}}, \bibinfo {author} {\bibfnamefont {J.}~\bibnamefont {{Haah}}}, \
  and\ \bibinfo {author} {\bibfnamefont {R.}~\bibnamefont {{Nandkishore}}},\
  }\href {\doibase 10.1103/PhysRevB.95.155133} {\bibfield  {journal} {\bibinfo
  {journal} {\prb}\ }\textbf {\bibinfo {volume} {95}},\ \bibinfo {eid} {155133}
  (\bibinfo {year} {2017})},\ \Eprint {http://arxiv.org/abs/1702.02952}
  {arXiv:1702.02952} \BibitemShut {NoStop}%
\bibitem [{\citenamefont {{Pai}}\ \emph {et~al.}(2019)\citenamefont {{Pai}},
  \citenamefont {{Pretko}},\ and\ \citenamefont {{Nandkishore}}}]{PaiCircuits}%
  \BibitemOpen
  \bibfield  {author} {\bibinfo {author} {\bibfnamefont {S.}~\bibnamefont
  {{Pai}}}, \bibinfo {author} {\bibfnamefont {M.}~\bibnamefont {{Pretko}}}, \
  and\ \bibinfo {author} {\bibfnamefont {R.~M.}\ \bibnamefont
  {{Nandkishore}}},\ }\href {\doibase 10.1103/PhysRevX.9.021003} {\bibfield
  {journal} {\bibinfo  {journal} {Physical Review X}\ }\textbf {\bibinfo
  {volume} {9}},\ \bibinfo {eid} {021003} (\bibinfo {year} {2019})},\ \Eprint
  {http://arxiv.org/abs/1807.09776} {arXiv:1807.09776} \BibitemShut {NoStop}%
\bibitem [{\citenamefont {{Gromov}}\ \emph {et~al.}(2020)\citenamefont
  {{Gromov}}, \citenamefont {{Lucas}},\ and\ \citenamefont
  {{Nandkishore}}}]{GromovHydrodynamics}%
  \BibitemOpen
  \bibfield  {author} {\bibinfo {author} {\bibfnamefont {A.}~\bibnamefont
  {{Gromov}}}, \bibinfo {author} {\bibfnamefont {A.}~\bibnamefont {{Lucas}}}, \
  and\ \bibinfo {author} {\bibfnamefont {R.~M.}\ \bibnamefont
  {{Nandkishore}}},\ }\href@noop {} {\  (\bibinfo {year} {2020})},\ \Eprint
  {http://arxiv.org/abs/2003.09429} {arXiv:2003.09429} \BibitemShut {NoStop}%
\bibitem [{\citenamefont {{Pai}}\ and\ \citenamefont
  {{Pretko}}(2019)}]{PaiScars}%
  \BibitemOpen
  \bibfield  {author} {\bibinfo {author} {\bibfnamefont {S.}~\bibnamefont
  {{Pai}}}\ and\ \bibinfo {author} {\bibfnamefont {M.}~\bibnamefont
  {{Pretko}}},\ }\href {\doibase 10.1103/PhysRevLett.123.136401} {\bibfield
  {journal} {\bibinfo  {journal} {\prl}\ }\textbf {\bibinfo {volume} {123}},\
  \bibinfo {eid} {136401} (\bibinfo {year} {2019})},\ \Eprint
  {http://arxiv.org/abs/1903.06173} {arXiv:1903.06173} \BibitemShut {NoStop}%
\bibitem [{\citenamefont {{He}}\ \emph {et~al.}(2020)\citenamefont {{He}},
  \citenamefont {{You}},\ and\ \citenamefont {{Prem}}}]{PremLSM}%
  \BibitemOpen
  \bibfield  {author} {\bibinfo {author} {\bibfnamefont {H.}~\bibnamefont
  {{He}}}, \bibinfo {author} {\bibfnamefont {Y.}~\bibnamefont {{You}}}, \ and\
  \bibinfo {author} {\bibfnamefont {A.}~\bibnamefont {{Prem}}},\ }\href
  {\doibase 10.1103/PhysRevB.101.165145} {\bibfield  {journal} {\bibinfo
  {journal} {\prb}\ }\textbf {\bibinfo {volume} {101}},\ \bibinfo {eid}
  {165145} (\bibinfo {year} {2020})},\ \Eprint
  {http://arxiv.org/abs/1912.10520} {arXiv:1912.10520} \BibitemShut {NoStop}%
\bibitem [{\citenamefont {{Dubinkin}}\ \emph {et~al.}(2020)\citenamefont
  {{Dubinkin}}, \citenamefont {{May-Mann}},\ and\ \citenamefont
  {{Hughes}}}]{HughesLSM}%
  \BibitemOpen
  \bibfield  {author} {\bibinfo {author} {\bibfnamefont {O.}~\bibnamefont
  {{Dubinkin}}}, \bibinfo {author} {\bibfnamefont {J.}~\bibnamefont
  {{May-Mann}}}, \ and\ \bibinfo {author} {\bibfnamefont {T.~L.}\ \bibnamefont
  {{Hughes}}},\ }\href@noop {} {\  (\bibinfo {year} {2020})},\ \Eprint
  {http://arxiv.org/abs/2001.04477} {arXiv:2001.04477} \BibitemShut {NoStop}%
\bibitem [{\citenamefont {{Shackleton}}\ and\ \citenamefont
  {{Scheurer}}(2020)}]{ScheurerPT}%
  \BibitemOpen
  \bibfield  {author} {\bibinfo {author} {\bibfnamefont {H.}~\bibnamefont
  {{Shackleton}}}\ and\ \bibinfo {author} {\bibfnamefont {M.~S.}\ \bibnamefont
  {{Scheurer}}},\ }\href {\doibase 10.1103/PhysRevResearch.2.033022} {\bibfield
   {journal} {\bibinfo  {journal} {Physical Review Research}\ }\textbf
  {\bibinfo {volume} {2}},\ \bibinfo {eid} {033022} (\bibinfo {year} {2020})},\
  \Eprint {http://arxiv.org/abs/2005.09668} {arXiv:2005.09668} \BibitemShut
  {NoStop}%
\bibitem [{\citenamefont {{Feldmeier}}\ \emph {et~al.}(2020)\citenamefont
  {{Feldmeier}}, \citenamefont {{Sala}}, \citenamefont {{De Tomasi}},
  \citenamefont {{Pollmann}},\ and\ \citenamefont {{Knap}}}]{PollmannMoment}%
  \BibitemOpen
  \bibfield  {author} {\bibinfo {author} {\bibfnamefont {J.}~\bibnamefont
  {{Feldmeier}}}, \bibinfo {author} {\bibfnamefont {P.}~\bibnamefont {{Sala}}},
  \bibinfo {author} {\bibfnamefont {G.}~\bibnamefont {{De Tomasi}}}, \bibinfo
  {author} {\bibfnamefont {F.}~\bibnamefont {{Pollmann}}}, \ and\ \bibinfo
  {author} {\bibfnamefont {M.}~\bibnamefont {{Knap}}},\ }\href {\doibase
  10.1103/PhysRevLett.125.245303} {\bibfield  {journal} {\bibinfo  {journal}
  {\prl}\ }\textbf {\bibinfo {volume} {125}},\ \bibinfo {eid} {245303}
  (\bibinfo {year} {2020})},\ \Eprint {http://arxiv.org/abs/2004.00635}
  {arXiv:2004.00635} \BibitemShut {NoStop}%
\bibitem [{\citenamefont {{Yuan}}\ \emph {et~al.}(2020)\citenamefont {{Yuan}},
  \citenamefont {{Chen}},\ and\ \citenamefont {{Ye}}}]{PengSuperfluid}%
  \BibitemOpen
  \bibfield  {author} {\bibinfo {author} {\bibfnamefont {J.-K.}\ \bibnamefont
  {{Yuan}}}, \bibinfo {author} {\bibfnamefont {S.~A.}\ \bibnamefont {{Chen}}},
  \ and\ \bibinfo {author} {\bibfnamefont {P.}~\bibnamefont {{Ye}}},\ }\href
  {\doibase 10.1103/PhysRevResearch.2.023267} {\bibfield  {journal} {\bibinfo
  {journal} {Physical Review Research}\ }\textbf {\bibinfo {volume} {2}},\
  \bibinfo {eid} {023267} (\bibinfo {year} {2020})},\ \Eprint
  {http://arxiv.org/abs/1911.02876} {arXiv:1911.02876} \BibitemShut {NoStop}%
\bibitem [{\citenamefont {{Yan}}(2019)}]{YanHolography}%
  \BibitemOpen
  \bibfield  {author} {\bibinfo {author} {\bibfnamefont {H.}~\bibnamefont
  {{Yan}}},\ }\href {\doibase 10.1103/PhysRevB.99.155126} {\bibfield  {journal}
  {\bibinfo  {journal} {\prb}\ }\textbf {\bibinfo {volume} {99}},\ \bibinfo
  {eid} {155126} (\bibinfo {year} {2019})},\ \Eprint
  {http://arxiv.org/abs/1807.05942} {arXiv:1807.05942} \BibitemShut {NoStop}%
\bibitem [{\citenamefont {{Yan}}(2020)}]{YanHolography2}%
  \BibitemOpen
  \bibfield  {author} {\bibinfo {author} {\bibfnamefont {H.}~\bibnamefont
  {{Yan}}},\ }\href {\doibase 10.1103/PhysRevB.102.161119} {\bibfield
  {journal} {\bibinfo  {journal} {\prb}\ }\textbf {\bibinfo {volume} {102}},\
  \bibinfo {eid} {161119} (\bibinfo {year} {2020})},\ \Eprint
  {http://arxiv.org/abs/1911.01007} {arXiv:1911.01007} \BibitemShut {NoStop}%
\bibitem [{\citenamefont {{Yan}}\ \emph {et~al.}(2020)\citenamefont {{Yan}},
  \citenamefont {{Benton}}, \citenamefont {{Jaubert}},\ and\ \citenamefont
  {{Shannon}}}]{YanBreathing}%
  \BibitemOpen
  \bibfield  {author} {\bibinfo {author} {\bibfnamefont {H.}~\bibnamefont
  {{Yan}}}, \bibinfo {author} {\bibfnamefont {O.}~\bibnamefont {{Benton}}},
  \bibinfo {author} {\bibfnamefont {L.~D.~C.}\ \bibnamefont {{Jaubert}}}, \
  and\ \bibinfo {author} {\bibfnamefont {N.}~\bibnamefont {{Shannon}}},\ }\href
  {\doibase 10.1103/PhysRevLett.124.127203} {\bibfield  {journal} {\bibinfo
  {journal} {\prl}\ }\textbf {\bibinfo {volume} {124}},\ \bibinfo {eid}
  {127203} (\bibinfo {year} {2020})},\ \Eprint
  {http://arxiv.org/abs/1902.10934} {arXiv:1902.10934} \BibitemShut {NoStop}%
\bibitem [{\citenamefont {{Pretko}}\ and\ \citenamefont
  {{Radzihovsky}}(2018)}]{PretkoElasticity}%
  \BibitemOpen
  \bibfield  {author} {\bibinfo {author} {\bibfnamefont {M.}~\bibnamefont
  {{Pretko}}}\ and\ \bibinfo {author} {\bibfnamefont {L.}~\bibnamefont
  {{Radzihovsky}}},\ }\href {\doibase 10.1103/PhysRevLett.120.195301}
  {\bibfield  {journal} {\bibinfo  {journal} {\prl}\ }\textbf {\bibinfo
  {volume} {120}},\ \bibinfo {eid} {195301} (\bibinfo {year} {2018})},\ \Eprint
  {http://arxiv.org/abs/1711.11044} {arXiv:1711.11044} \BibitemShut {NoStop}%
\bibitem [{\citenamefont {{Pretko}}\ \emph {et~al.}(2019)\citenamefont
  {{Pretko}}, \citenamefont {{Zhai}},\ and\ \citenamefont
  {{Radzihovsky}}}]{PretkoDuality}%
  \BibitemOpen
  \bibfield  {author} {\bibinfo {author} {\bibfnamefont {M.}~\bibnamefont
  {{Pretko}}}, \bibinfo {author} {\bibfnamefont {Z.}~\bibnamefont {{Zhai}}}, \
  and\ \bibinfo {author} {\bibfnamefont {L.}~\bibnamefont {{Radzihovsky}}},\
  }\href {\doibase 10.1103/PhysRevB.100.134113} {\bibfield  {journal} {\bibinfo
   {journal} {\prb}\ }\textbf {\bibinfo {volume} {100}},\ \bibinfo {eid}
  {134113} (\bibinfo {year} {2019})},\ \Eprint
  {http://arxiv.org/abs/1907.12577} {arXiv:1907.12577} \BibitemShut {NoStop}%
\bibitem [{\citenamefont {{Hal{\'a}sz}}\ \emph {et~al.}(2017)\citenamefont
  {{Hal{\'a}sz}}, \citenamefont {{Hsieh}},\ and\ \citenamefont
  {{Balents}}}]{HsiehChains}%
  \BibitemOpen
  \bibfield  {author} {\bibinfo {author} {\bibfnamefont {G.~B.}\ \bibnamefont
  {{Hal{\'a}sz}}}, \bibinfo {author} {\bibfnamefont {T.~H.}\ \bibnamefont
  {{Hsieh}}}, \ and\ \bibinfo {author} {\bibfnamefont {L.}~\bibnamefont
  {{Balents}}},\ }\href {\doibase 10.1103/PhysRevLett.119.257202} {\bibfield
  {journal} {\bibinfo  {journal} {\prl}\ }\textbf {\bibinfo {volume} {119}},\
  \bibinfo {eid} {257202} (\bibinfo {year} {2017})},\ \Eprint
  {http://arxiv.org/abs/1707.02308} {arXiv:1707.02308} \BibitemShut {NoStop}%
\bibitem [{\citenamefont {{Fuji}}(2019)}]{FujiLayer}%
  \BibitemOpen
  \bibfield  {author} {\bibinfo {author} {\bibfnamefont {Y.}~\bibnamefont
  {{Fuji}}},\ }\href {\doibase 10.1103/PhysRevB.100.235115} {\bibfield
  {journal} {\bibinfo  {journal} {\prb}\ }\textbf {\bibinfo {volume} {100}},\
  \bibinfo {eid} {235115} (\bibinfo {year} {2019})},\ \Eprint
  {http://arxiv.org/abs/1908.02257} {arXiv:1908.02257} \BibitemShut {NoStop}%
\bibitem [{\citenamefont {{Slagle}}\ and\ \citenamefont
  {{Kim}}(2017{\natexlab{a}})}]{Slagle2spin}%
  \BibitemOpen
  \bibfield  {author} {\bibinfo {author} {\bibfnamefont {K.}~\bibnamefont
  {{Slagle}}}\ and\ \bibinfo {author} {\bibfnamefont {Y.~B.}\ \bibnamefont
  {{Kim}}},\ }\href {\doibase 10.1103/PhysRevB.96.165106} {\bibfield  {journal}
  {\bibinfo  {journal} {\prb}\ }\textbf {\bibinfo {volume} {96}},\ \bibinfo
  {eid} {165106} (\bibinfo {year} {2017}{\natexlab{a}})},\ \Eprint
  {http://arxiv.org/abs/1704.03870} {arXiv:1704.03870} \BibitemShut {NoStop}%
\bibitem [{\citenamefont {{Prem}}\ \emph {et~al.}(2018)\citenamefont {{Prem}},
  \citenamefont {{Vijay}}, \citenamefont {{Chou}}, \citenamefont {{Pretko}},\
  and\ \citenamefont {{Nandkishore}}}]{PretkoPinch}%
  \BibitemOpen
  \bibfield  {author} {\bibinfo {author} {\bibfnamefont {A.}~\bibnamefont
  {{Prem}}}, \bibinfo {author} {\bibfnamefont {S.}~\bibnamefont {{Vijay}}},
  \bibinfo {author} {\bibfnamefont {Y.-Z.}\ \bibnamefont {{Chou}}}, \bibinfo
  {author} {\bibfnamefont {M.}~\bibnamefont {{Pretko}}}, \ and\ \bibinfo
  {author} {\bibfnamefont {R.~M.}\ \bibnamefont {{Nandkishore}}},\ }\href
  {\doibase 10.1103/PhysRevB.98.165140} {\bibfield  {journal} {\bibinfo
  {journal} {\prb}\ }\textbf {\bibinfo {volume} {98}},\ \bibinfo {eid} {165140}
  (\bibinfo {year} {2018})},\ \Eprint {http://arxiv.org/abs/1806.04148}
  {arXiv:1806.04148} \BibitemShut {NoStop}%
\bibitem [{\citenamefont {{Nguyen}}\ \emph {et~al.}(2020)\citenamefont
  {{Nguyen}}, \citenamefont {{Gromov}},\ and\ \citenamefont
  {{Moroz}}}]{GromovDuality}%
  \BibitemOpen
  \bibfield  {author} {\bibinfo {author} {\bibfnamefont {D.}~\bibnamefont
  {{Nguyen}}}, \bibinfo {author} {\bibfnamefont {A.}~\bibnamefont {{Gromov}}},
  \ and\ \bibinfo {author} {\bibfnamefont {S.}~\bibnamefont {{Moroz}}},\ }\href
  {\doibase 10.21468/SciPostPhys.9.5.076} {\bibfield  {journal} {\bibinfo
  {journal} {SciPost Physics}\ }\textbf {\bibinfo {volume} {9}},\ \bibinfo
  {eid} {076} (\bibinfo {year} {2020})},\ \Eprint
  {http://arxiv.org/abs/2005.12317} {arXiv:2005.12317} \BibitemShut {NoStop}%
\bibitem [{\citenamefont {{Doshi}}\ and\ \citenamefont
  {{Gromov}}(2020)}]{GromovVortices}%
  \BibitemOpen
  \bibfield  {author} {\bibinfo {author} {\bibfnamefont {D.}~\bibnamefont
  {{Doshi}}}\ and\ \bibinfo {author} {\bibfnamefont {A.}~\bibnamefont
  {{Gromov}}},\ }\href@noop {} {\  (\bibinfo {year} {2020})},\ \Eprint
  {http://arxiv.org/abs/2005.03015} {arXiv:2005.03015} \BibitemShut {NoStop}%
\bibitem [{\citenamefont {{Pankov}}\ \emph {et~al.}(2007)\citenamefont
  {{Pankov}}, \citenamefont {{Moessner}},\ and\ \citenamefont
  {{Sondhi}}}]{SondhiPlaquettes}%
  \BibitemOpen
  \bibfield  {author} {\bibinfo {author} {\bibfnamefont {S.}~\bibnamefont
  {{Pankov}}}, \bibinfo {author} {\bibfnamefont {R.}~\bibnamefont
  {{Moessner}}}, \ and\ \bibinfo {author} {\bibfnamefont {S.~L.}\ \bibnamefont
  {{Sondhi}}},\ }\href {\doibase 10.1103/PhysRevB.76.104436} {\bibfield
  {journal} {\bibinfo  {journal} {\prb}\ }\textbf {\bibinfo {volume} {76}},\
  \bibinfo {eid} {104436} (\bibinfo {year} {2007})},\ \Eprint
  {http://arxiv.org/abs/0705.0846} {arXiv:0705.0846} \BibitemShut {NoStop}%
\bibitem [{\citenamefont {{Xu}}\ and\ \citenamefont
  {{Wu}}(2008)}]{CenkePlaquettes}%
  \BibitemOpen
  \bibfield  {author} {\bibinfo {author} {\bibfnamefont {C.}~\bibnamefont
  {{Xu}}}\ and\ \bibinfo {author} {\bibfnamefont {C.}~\bibnamefont {{Wu}}},\
  }\href {\doibase 10.1103/PhysRevB.77.134449} {\bibfield  {journal} {\bibinfo
  {journal} {\prb}\ }\textbf {\bibinfo {volume} {77}},\ \bibinfo {eid} {134449}
  (\bibinfo {year} {2008})},\ \Eprint {http://arxiv.org/abs/0801.0744}
  {arXiv:0801.0744} \BibitemShut {NoStop}%
\bibitem [{\citenamefont {Sous}\ and\ \citenamefont
  {Pretko}(2020)}]{SousPolarons}%
  \BibitemOpen
  \bibfield  {author} {\bibinfo {author} {\bibfnamefont {J.}~\bibnamefont
  {Sous}}\ and\ \bibinfo {author} {\bibfnamefont {M.}~\bibnamefont {Pretko}},\
  }\href {\doibase 10.1103/PhysRevB.102.214437} {\bibfield  {journal} {\bibinfo
   {journal} {Phys. Rev. B}\ }\textbf {\bibinfo {volume} {102}},\ \bibinfo
  {pages} {214437} (\bibinfo {year} {2020})},\ \Eprint
  {http://arxiv.org/abs/1904.08424} {arXiv:1904.08424} \BibitemShut {NoStop}%
\bibitem [{\citenamefont {You}\ and\ \citenamefont {von
  Oppen}(2019)}]{YizhiLego}%
  \BibitemOpen
  \bibfield  {author} {\bibinfo {author} {\bibfnamefont {Y.}~\bibnamefont
  {You}}\ and\ \bibinfo {author} {\bibfnamefont {F.}~\bibnamefont {von
  Oppen}},\ }\href {\doibase 10.1103/PhysRevResearch.1.013011} {\bibfield
  {journal} {\bibinfo  {journal} {Phys. Rev. Research}\ }\textbf {\bibinfo
  {volume} {1}},\ \bibinfo {pages} {013011} (\bibinfo {year} {2019})},\ \Eprint
  {http://arxiv.org/abs/1812.06091} {arXiv:1812.06091} \BibitemShut {NoStop}%
\bibitem [{\citenamefont {{Pretko}}(2017{\natexlab{a}})}]{PretkoGravity}%
  \BibitemOpen
  \bibfield  {author} {\bibinfo {author} {\bibfnamefont {M.}~\bibnamefont
  {{Pretko}}},\ }\href {\doibase 10.1103/PhysRevD.96.024051} {\bibfield
  {journal} {\bibinfo  {journal} {\prd}\ }\textbf {\bibinfo {volume} {96}},\
  \bibinfo {eid} {024051} (\bibinfo {year} {2017}{\natexlab{a}})},\ \Eprint
  {http://arxiv.org/abs/1702.07613} {arXiv:1702.07613} \BibitemShut {NoStop}%
\bibitem [{\citenamefont {{Pretko}}(2017{\natexlab{b}})}]{PretkoU1}%
  \BibitemOpen
  \bibfield  {author} {\bibinfo {author} {\bibfnamefont {M.}~\bibnamefont
  {{Pretko}}},\ }\href {\doibase 10.1103/PhysRevB.95.115139} {\bibfield
  {journal} {\bibinfo  {journal} {\prb}\ }\textbf {\bibinfo {volume} {95}},\
  \bibinfo {eid} {115139} (\bibinfo {year} {2017}{\natexlab{b}})},\ \Eprint
  {http://arxiv.org/abs/1604.05329} {arXiv:1604.05329} \BibitemShut {NoStop}%
\bibitem [{\citenamefont {{Rasmussen}}\ \emph {et~al.}(2016)\citenamefont
  {{Rasmussen}}, \citenamefont {{You}},\ and\ \citenamefont
  {{Xu}}}]{RasmussenU1}%
  \BibitemOpen
  \bibfield  {author} {\bibinfo {author} {\bibfnamefont {A.}~\bibnamefont
  {{Rasmussen}}}, \bibinfo {author} {\bibfnamefont {Y.-Z.}\ \bibnamefont
  {{You}}}, \ and\ \bibinfo {author} {\bibfnamefont {C.}~\bibnamefont {{Xu}}},\
  }\href@noop {} {\  (\bibinfo {year} {2016})},\ \Eprint
  {http://arxiv.org/abs/1601.08235} {arXiv:1601.08235} \BibitemShut {NoStop}%
\bibitem [{\citenamefont {{Pretko}}(2017{\natexlab{c}})}]{PretkoEM}%
  \BibitemOpen
  \bibfield  {author} {\bibinfo {author} {\bibfnamefont {M.}~\bibnamefont
  {{Pretko}}},\ }\href {\doibase 10.1103/PhysRevB.96.035119} {\bibfield
  {journal} {\bibinfo  {journal} {\prb}\ }\textbf {\bibinfo {volume} {96}},\
  \bibinfo {eid} {035119} (\bibinfo {year} {2017}{\natexlab{c}})},\ \Eprint
  {http://arxiv.org/abs/1606.08857} {arXiv:1606.08857} \BibitemShut {NoStop}%
\bibitem [{\citenamefont {{Radzihovsky}}\ and\ \citenamefont
  {{Hermele}}(2020)}]{RadzihovskyVector}%
  \BibitemOpen
  \bibfield  {author} {\bibinfo {author} {\bibfnamefont {L.}~\bibnamefont
  {{Radzihovsky}}}\ and\ \bibinfo {author} {\bibfnamefont {M.}~\bibnamefont
  {{Hermele}}},\ }\href {\doibase 10.1103/PhysRevLett.124.050402} {\bibfield
  {journal} {\bibinfo  {journal} {\prl}\ }\textbf {\bibinfo {volume} {124}},\
  \bibinfo {eid} {050402} (\bibinfo {year} {2020})},\ \Eprint
  {http://arxiv.org/abs/1905.06951} {arXiv:1905.06951} \BibitemShut {NoStop}%
\bibitem [{\citenamefont {Seiberg}\ and\ \citenamefont
  {Shao}(2020)}]{SeibergU1}%
  \BibitemOpen
  \bibfield  {author} {\bibinfo {author} {\bibfnamefont {N.}~\bibnamefont
  {Seiberg}}\ and\ \bibinfo {author} {\bibfnamefont {S.-H.}\ \bibnamefont
  {Shao}},\ }\href {\doibase 10.21468/SciPostPhys.9.4.046} {\bibfield
  {journal} {\bibinfo  {journal} {SciPost Phys.}\ }\textbf {\bibinfo {volume}
  {9}},\ \bibinfo {pages} {46} (\bibinfo {year} {2020})},\ \Eprint
  {http://arxiv.org/abs/2004.00015} {arXiv:2004.00015} \BibitemShut {NoStop}%
\bibitem [{\citenamefont {{Seiberg}}(2020)}]{SeibergSymmetry}%
  \BibitemOpen
  \bibfield  {author} {\bibinfo {author} {\bibfnamefont {N.}~\bibnamefont
  {{Seiberg}}},\ }\href {\doibase 10.21468/SciPostPhys.8.4.050} {\bibfield
  {journal} {\bibinfo  {journal} {SciPost Physics}\ }\textbf {\bibinfo {volume}
  {8}},\ \bibinfo {eid} {050} (\bibinfo {year} {2020})},\ \Eprint
  {http://arxiv.org/abs/1909.10544} {arXiv:1909.10544} \BibitemShut {NoStop}%
\bibitem [{\citenamefont {{Pretko}}(2017{\natexlab{d}})}]{PretkoTemperature}%
  \BibitemOpen
  \bibfield  {author} {\bibinfo {author} {\bibfnamefont {M.}~\bibnamefont
  {{Pretko}}},\ }\href {\doibase 10.1103/PhysRevB.96.115102} {\bibfield
  {journal} {\bibinfo  {journal} {\prb}\ }\textbf {\bibinfo {volume} {96}},\
  \bibinfo {eid} {115102} (\bibinfo {year} {2017}{\natexlab{d}})},\ \Eprint
  {http://arxiv.org/abs/1706.01899} {arXiv:1706.01899} \BibitemShut {NoStop}%
\bibitem [{\citenamefont {{Griffin}}\ \emph {et~al.}(2015)\citenamefont
  {{Griffin}}, \citenamefont {{Grosvenor}}, \citenamefont {{Ho{\v{r}}ava}},\
  and\ \citenamefont {{Yan}}}]{PolyShift}%
  \BibitemOpen
  \bibfield  {author} {\bibinfo {author} {\bibfnamefont {T.}~\bibnamefont
  {{Griffin}}}, \bibinfo {author} {\bibfnamefont {K.~T.}\ \bibnamefont
  {{Grosvenor}}}, \bibinfo {author} {\bibfnamefont {P.}~\bibnamefont
  {{Ho{\v{r}}ava}}}, \ and\ \bibinfo {author} {\bibfnamefont {Z.}~\bibnamefont
  {{Yan}}},\ }\href {\doibase 10.1007/s00220-015-2461-2} {\bibfield  {journal}
  {\bibinfo  {journal} {Communications in Mathematical Physics}\ }\textbf
  {\bibinfo {volume} {340}},\ \bibinfo {pages} {985} (\bibinfo {year}
  {2015})},\ \Eprint {http://arxiv.org/abs/1412.1046} {arXiv:1412.1046}
  \BibitemShut {NoStop}%
\bibitem [{Ifo()}]{Ifoot:fractal}%
  \BibitemOpen
  \href@noop {} {}\bibinfo {note} {See
  \refscite{HaahCode,YoshidaCode,ZhenghanHaah,VijayXcube} for gapped fracton
  models with fractal operators.}\BibitemShut {Stop}%
\bibitem [{\citenamefont {{Shirley}}\ \emph {et~al.}(2019)\citenamefont
  {{Shirley}}, \citenamefont {{Slagle}},\ and\ \citenamefont
  {{Chen}}}]{gauging}%
  \BibitemOpen
  \bibfield  {author} {\bibinfo {author} {\bibfnamefont {W.}~\bibnamefont
  {{Shirley}}}, \bibinfo {author} {\bibfnamefont {K.}~\bibnamefont {{Slagle}}},
  \ and\ \bibinfo {author} {\bibfnamefont {X.}~\bibnamefont {{Chen}}},\ }\href
  {\doibase 10.21468/SciPostPhys.6.4.041} {\bibfield  {journal} {\bibinfo
  {journal} {SciPost Physics}\ }\textbf {\bibinfo {volume} {6}},\ \bibinfo
  {eid} {041} (\bibinfo {year} {2019})},\ \Eprint
  {http://arxiv.org/abs/1806.08679} {arXiv:1806.08679} \BibitemShut {NoStop}%
\bibitem [{\citenamefont {Shirley}\ \emph {et~al.}(2019)\citenamefont
  {Shirley}, \citenamefont {Slagle},\ and\ \citenamefont
  {Chen}}]{ShirleyEntanglement}%
  \BibitemOpen
  \bibfield  {author} {\bibinfo {author} {\bibfnamefont {W.}~\bibnamefont
  {Shirley}}, \bibinfo {author} {\bibfnamefont {K.}~\bibnamefont {Slagle}}, \
  and\ \bibinfo {author} {\bibfnamefont {X.}~\bibnamefont {Chen}},\ }\href
  {\doibase 10.21468/SciPostPhys.6.1.015} {\bibfield  {journal} {\bibinfo
  {journal} {SciPost Phys.}\ }\textbf {\bibinfo {volume} {6}},\ \bibinfo
  {pages} {15} (\bibinfo {year} {2019})},\ \Eprint
  {http://arxiv.org/abs/1803.10426} {arXiv:1803.10426} \BibitemShut {NoStop}%
\bibitem [{\citenamefont {{Pai}}\ and\ \citenamefont
  {{Hermele}}(2019)}]{PaiFusion}%
  \BibitemOpen
  \bibfield  {author} {\bibinfo {author} {\bibfnamefont {S.}~\bibnamefont
  {{Pai}}}\ and\ \bibinfo {author} {\bibfnamefont {M.}~\bibnamefont
  {{Hermele}}},\ }\href {\doibase 10.1103/PhysRevB.100.195136} {\bibfield
  {journal} {\bibinfo  {journal} {\prb}\ }\textbf {\bibinfo {volume} {100}},\
  \bibinfo {eid} {195136} (\bibinfo {year} {2019})},\ \Eprint
  {http://arxiv.org/abs/1903.11625} {arXiv:1903.11625} \BibitemShut {NoStop}%
\bibitem [{\citenamefont {{Shirley}}\ \emph {et~al.}(2018)\citenamefont
  {{Shirley}}, \citenamefont {{Slagle}}, \citenamefont {{Wang}},\ and\
  \citenamefont {{Chen}}}]{3manifolds}%
  \BibitemOpen
  \bibfield  {author} {\bibinfo {author} {\bibfnamefont {W.}~\bibnamefont
  {{Shirley}}}, \bibinfo {author} {\bibfnamefont {K.}~\bibnamefont {{Slagle}}},
  \bibinfo {author} {\bibfnamefont {Z.}~\bibnamefont {{Wang}}}, \ and\ \bibinfo
  {author} {\bibfnamefont {X.}~\bibnamefont {{Chen}}},\ }\href {\doibase
  10.1103/PhysRevX.8.031051} {\bibfield  {journal} {\bibinfo  {journal}
  {Physical Review X}\ }\textbf {\bibinfo {volume} {8}},\ \bibinfo {eid}
  {031051} (\bibinfo {year} {2018})},\ \Eprint
  {http://arxiv.org/abs/1712.05892} {arXiv:1712.05892} \BibitemShut {NoStop}%
\bibitem [{\citenamefont {Aasen}\ \emph {et~al.}(2020)\citenamefont {Aasen},
  \citenamefont {Bulmash}, \citenamefont {Prem}, \citenamefont {Slagle},\ and\
  \citenamefont {Williamson}}]{defectNetworks}%
  \BibitemOpen
  \bibfield  {author} {\bibinfo {author} {\bibfnamefont {D.}~\bibnamefont
  {Aasen}}, \bibinfo {author} {\bibfnamefont {D.}~\bibnamefont {Bulmash}},
  \bibinfo {author} {\bibfnamefont {A.}~\bibnamefont {Prem}}, \bibinfo {author}
  {\bibfnamefont {K.}~\bibnamefont {Slagle}}, \ and\ \bibinfo {author}
  {\bibfnamefont {D.~J.}\ \bibnamefont {Williamson}},\ }\href {\doibase
  10.1103/PhysRevResearch.2.043165} {\bibfield  {journal} {\bibinfo  {journal}
  {Phys. Rev. Research}\ }\textbf {\bibinfo {volume} {2}},\ \bibinfo {pages}
  {043165} (\bibinfo {year} {2020})},\ \Eprint
  {http://arxiv.org/abs/2002.05166} {arXiv:2002.05166} \BibitemShut {NoStop}%
\bibitem [{\citenamefont {Wen}(2020)}]{WenCellular}%
  \BibitemOpen
  \bibfield  {author} {\bibinfo {author} {\bibfnamefont {X.-G.}\ \bibnamefont
  {Wen}},\ }\href {\doibase 10.1103/PhysRevResearch.2.033300} {\bibfield
  {journal} {\bibinfo  {journal} {Phys. Rev. Research}\ }\textbf {\bibinfo
  {volume} {2}},\ \bibinfo {pages} {033300} (\bibinfo {year} {2020})},\ \Eprint
  {http://arxiv.org/abs/2002.02433} {arXiv:2002.02433} \BibitemShut {NoStop}%
\bibitem [{\citenamefont {{Wang}}(2020)}]{JuvenCellular}%
  \BibitemOpen
  \bibfield  {author} {\bibinfo {author} {\bibfnamefont {J.}~\bibnamefont
  {{Wang}}},\ }\href@noop {} {\  (\bibinfo {year} {2020})},\ \Eprint
  {http://arxiv.org/abs/2002.12932} {arXiv:2002.12932} \BibitemShut {NoStop}%
\bibitem [{\citenamefont {{Slagle}}\ \emph
  {et~al.}(2019{\natexlab{a}})\citenamefont {{Slagle}}, \citenamefont
  {{Aasen}},\ and\ \citenamefont {{Williamson}}}]{stringMembraneNet}%
  \BibitemOpen
  \bibfield  {author} {\bibinfo {author} {\bibfnamefont {K.}~\bibnamefont
  {{Slagle}}}, \bibinfo {author} {\bibfnamefont {D.}~\bibnamefont {{Aasen}}}, \
  and\ \bibinfo {author} {\bibfnamefont {D.}~\bibnamefont {{Williamson}}},\
  }\href {\doibase 10.21468/SciPostPhys.6.4.043} {\bibfield  {journal}
  {\bibinfo  {journal} {SciPost Physics}\ }\textbf {\bibinfo {volume} {6}},\
  \bibinfo {eid} {043} (\bibinfo {year} {2019}{\natexlab{a}})},\ \Eprint
  {http://arxiv.org/abs/1812.01613} {arXiv:1812.01613} \BibitemShut {NoStop}%
\bibitem [{\citenamefont {Slagle}\ and\ \citenamefont
  {Kim}(2018)}]{SlagleLattices}%
  \BibitemOpen
  \bibfield  {author} {\bibinfo {author} {\bibfnamefont {K.}~\bibnamefont
  {Slagle}}\ and\ \bibinfo {author} {\bibfnamefont {Y.~B.}\ \bibnamefont
  {Kim}},\ }\href {\doibase 10.1103/PhysRevB.97.165106} {\bibfield  {journal}
  {\bibinfo  {journal} {Phys. Rev. B}\ }\textbf {\bibinfo {volume} {97}},\
  \bibinfo {pages} {165106} (\bibinfo {year} {2018})},\ \Eprint
  {http://arxiv.org/abs/1712.04511} {arXiv:1712.04511} \BibitemShut {NoStop}%
\bibitem [{\citenamefont {{Slagle}}\ and\ \citenamefont
  {{Kim}}(2017{\natexlab{b}})}]{XcubeQFT}%
  \BibitemOpen
  \bibfield  {author} {\bibinfo {author} {\bibfnamefont {K.}~\bibnamefont
  {{Slagle}}}\ and\ \bibinfo {author} {\bibfnamefont {Y.~B.}\ \bibnamefont
  {{Kim}}},\ }\href {\doibase 10.1103/PhysRevB.96.195139} {\bibfield  {journal}
  {\bibinfo  {journal} {\prb}\ }\textbf {\bibinfo {volume} {96}},\ \bibinfo
  {eid} {195139} (\bibinfo {year} {2017}{\natexlab{b}})},\ \Eprint
  {http://arxiv.org/abs/1708.04619} {arXiv:1708.04619} \BibitemShut {NoStop}%
\bibitem [{\citenamefont {Seiberg}\ and\ \citenamefont
  {Shao}(2021)}]{Seiberg3}%
  \BibitemOpen
  \bibfield  {author} {\bibinfo {author} {\bibfnamefont {N.}~\bibnamefont
  {Seiberg}}\ and\ \bibinfo {author} {\bibfnamefont {S.-H.}\ \bibnamefont
  {Shao}},\ }\href {\doibase 10.21468/SciPostPhys.10.1.003} {\bibfield
  {journal} {\bibinfo  {journal} {SciPost Phys.}\ }\textbf {\bibinfo {volume}
  {10}},\ \bibinfo {pages} {3} (\bibinfo {year} {2021})},\ \Eprint
  {http://arxiv.org/abs/2004.06115} {arXiv:2004.06115} \BibitemShut {NoStop}%
\bibitem [{\citenamefont {Gorantla}\ \emph {et~al.}(2020)\citenamefont
  {Gorantla}, \citenamefont {Lam}, \citenamefont {Seiberg},\ and\ \citenamefont
  {Shao}}]{Seiberg4}%
  \BibitemOpen
  \bibfield  {author} {\bibinfo {author} {\bibfnamefont {P.}~\bibnamefont
  {Gorantla}}, \bibinfo {author} {\bibfnamefont {H.~T.}\ \bibnamefont {Lam}},
  \bibinfo {author} {\bibfnamefont {N.}~\bibnamefont {Seiberg}}, \ and\
  \bibinfo {author} {\bibfnamefont {S.-H.}\ \bibnamefont {Shao}},\ }\href
  {\doibase 10.21468/SciPostPhys.9.5.073} {\bibfield  {journal} {\bibinfo
  {journal} {SciPost Phys.}\ }\textbf {\bibinfo {volume} {9}},\ \bibinfo
  {pages} {73} (\bibinfo {year} {2020})},\ \Eprint
  {http://arxiv.org/abs/2007.04904} {arXiv:2007.04904} \BibitemShut {NoStop}%
\bibitem [{\citenamefont {{Fontana}}\ \emph {et~al.}(2020)\citenamefont
  {{Fontana}}, \citenamefont {{Gomes}},\ and\ \citenamefont
  {{Chamon}}}]{ChamonCS}%
  \BibitemOpen
  \bibfield  {author} {\bibinfo {author} {\bibfnamefont {W.~B.}\ \bibnamefont
  {{Fontana}}}, \bibinfo {author} {\bibfnamefont {P.~R.~S.}\ \bibnamefont
  {{Gomes}}}, \ and\ \bibinfo {author} {\bibfnamefont {C.}~\bibnamefont
  {{Chamon}}},\ }\href@noop {} {\  (\bibinfo {year} {2020})},\ \Eprint
  {http://arxiv.org/abs/2006.10071} {arXiv:2006.10071} \BibitemShut {NoStop}%
\bibitem [{\citenamefont {{You}}\ \emph
  {et~al.}(2020{\natexlab{a}})\citenamefont {{You}}, \citenamefont {{Devakul}},
  \citenamefont {{Sondhi}},\ and\ \citenamefont {{Burnell}}}]{YizhiBF}%
  \BibitemOpen
  \bibfield  {author} {\bibinfo {author} {\bibfnamefont {Y.}~\bibnamefont
  {{You}}}, \bibinfo {author} {\bibfnamefont {T.}~\bibnamefont {{Devakul}}},
  \bibinfo {author} {\bibfnamefont {S.~L.}\ \bibnamefont {{Sondhi}}}, \ and\
  \bibinfo {author} {\bibfnamefont {F.~J.}\ \bibnamefont {{Burnell}}},\ }\href
  {\doibase 10.1103/PhysRevResearch.2.023249} {\bibfield  {journal} {\bibinfo
  {journal} {Physical Review Research}\ }\textbf {\bibinfo {volume} {2}},\
  \bibinfo {eid} {023249} (\bibinfo {year} {2020}{\natexlab{a}})},\ \Eprint
  {http://arxiv.org/abs/1904.11530} {arXiv:1904.11530} \BibitemShut {NoStop}%
\bibitem [{\citenamefont {{You}}\ \emph
  {et~al.}(2020{\natexlab{b}})\citenamefont {{You}}, \citenamefont {{Devakul}},
  \citenamefont {{Burnell}},\ and\ \citenamefont {{Sondhi}}}]{YizhiChernSimon}%
  \BibitemOpen
  \bibfield  {author} {\bibinfo {author} {\bibfnamefont {Y.}~\bibnamefont
  {{You}}}, \bibinfo {author} {\bibfnamefont {T.}~\bibnamefont {{Devakul}}},
  \bibinfo {author} {\bibfnamefont {F.~J.}\ \bibnamefont {{Burnell}}}, \ and\
  \bibinfo {author} {\bibfnamefont {S.~L.}\ \bibnamefont {{Sondhi}}},\ }\href
  {\doibase 10.1016/j.aop.2020.168140} {\bibfield  {journal} {\bibinfo
  {journal} {Annals of Physics}\ }\textbf {\bibinfo {volume} {416}},\ \bibinfo
  {eid} {168140} (\bibinfo {year} {2020}{\natexlab{b}})},\ \Eprint
  {http://arxiv.org/abs/1805.09800} {arXiv:1805.09800} \BibitemShut {NoStop}%
\bibitem [{Cfo()}]{Cfoot:foliatedOrder}%
  \BibitemOpen
  \href@noop {} {}\bibinfo {note} {See \appref{app:RG}{F} and Section 2 of
  \refcite{ShirleyEntanglement} for more details on foliated fracton
  phases.}\BibitemShut {Stop}%
\bibitem [{\citenamefont {{Shirley}}\ \emph {et~al.}(2020)\citenamefont
  {{Shirley}}, \citenamefont {{Slagle}},\ and\ \citenamefont
  {{Chen}}}]{twisted}%
  \BibitemOpen
  \bibfield  {author} {\bibinfo {author} {\bibfnamefont {W.}~\bibnamefont
  {{Shirley}}}, \bibinfo {author} {\bibfnamefont {K.}~\bibnamefont {{Slagle}}},
  \ and\ \bibinfo {author} {\bibfnamefont {X.}~\bibnamefont {{Chen}}},\ }\href
  {\doibase 10.1103/PhysRevB.102.115103} {\bibfield  {journal} {\bibinfo
  {journal} {\prb}\ }\textbf {\bibinfo {volume} {102}},\ \bibinfo {eid}
  {115103} (\bibinfo {year} {2020})},\ \Eprint
  {http://arxiv.org/abs/1907.09048} {arXiv:1907.09048} \BibitemShut {NoStop}%
\bibitem [{\citenamefont {Slagle}()}]{talk}%
  \BibitemOpen
  \bibfield  {author} {\bibinfo {author} {\bibfnamefont {K.}~\bibnamefont
  {Slagle}},\ }\href
  {https://harvard.zoom.us/rec/share/6JV4dq_J7FNJGYXjwV7tAbAzAaa9eaa8gHcW_qUNxUt6lliFajvuU8tFGgT4mLs9}
  {\enquote {\bibinfo {title} {{Foliated QFT and Topological Defect Networks of
  Fracton Order}},}\ }\bibinfo {note} {Harvard CMSA, June 18
  (2020)}\BibitemShut {NoStop}%
\bibitem [{\citenamefont {Frankel}(2011)}]{FrankelGeometry}%
  \BibitemOpen
  \bibfield  {author} {\bibinfo {author} {\bibfnamefont {T.}~\bibnamefont
  {Frankel}},\ }\href@noop {} {\emph {\bibinfo {title} {The Geometry of
  Physics: An Introduction}}}\ (\bibinfo  {publisher} {Cambridge University
  Press},\ \bibinfo {year} {2011})\BibitemShut {NoStop}%
\bibitem [{\citenamefont {{Godbillon}}\ and\ \citenamefont
  {{Vey}}(1971)}]{GodbillonVey}%
  \BibitemOpen
  \bibfield  {author} {\bibinfo {author} {\bibfnamefont {C.}~\bibnamefont
  {{Godbillon}}}\ and\ \bibinfo {author} {\bibfnamefont {J.}~\bibnamefont
  {{Vey}}},\ }\href@noop {} {\bibfield  {journal} {\bibinfo  {journal} {C. R.
  Acad. Sci. Paris}\ }\textbf {\bibinfo {volume} {273}},\ \bibinfo {pages} {92}
  (\bibinfo {year} {1971})}\BibitemShut {NoStop}%
\bibitem [{\citenamefont {{Kotschick}}(2001)}]{GodbillonVeyFamilies}%
  \BibitemOpen
  \bibfield  {author} {\bibinfo {author} {\bibfnamefont {D.}~\bibnamefont
  {{Kotschick}}},\ }\href@noop {} {\  (\bibinfo {year} {2001})},\ \Eprint
  {http://arxiv.org/abs/math/0111137} {arXiv:math/0111137} \BibitemShut
  {NoStop}%
\bibitem [{Dfo()}]{Dfoot:BF}%
  \BibitemOpen
  \href@noop {} {}\bibinfo {note} {See Appendix A and B of \refcite{XcubeQFT}
  for a review of the connection between BF theory and toric code, and see
  Section 2.2 of \refcite{SeibergBFZn} for the equivalence to $Z_N$ gauge
  theory.}\BibitemShut {Stop}%
\bibitem [{\citenamefont {{Kitaev}}(2003)}]{KitaevAnyon}%
  \BibitemOpen
  \bibfield  {author} {\bibinfo {author} {\bibfnamefont {A.~Y.}\ \bibnamefont
  {{Kitaev}}},\ }\href {\doibase 10.1016/S0003-4916(02)00018-0} {\bibfield
  {journal} {\bibinfo  {journal} {Annals of Physics}\ }\textbf {\bibinfo
  {volume} {303}},\ \bibinfo {pages} {2} (\bibinfo {year} {2003})},\ \Eprint
  {http://arxiv.org/abs/quant-ph/9707021} {arXiv:quant-ph/9707021} \BibitemShut
  {NoStop}%
\bibitem [{Jfo()}]{Jfoot:XcubeFoliation}%
  \BibitemOpen
  \href@noop {} {}\bibinfo {note} {See e.g. Fig. 3(b) of
  \refcite{SlagleLattices} for a 4-foliated X-cube model where $e^1 = +
  \frac{1}{2} dx - \frac{\sqrt{3}}{2} dy$, $e^2 = - \frac{1}{2} dx -
  \frac{\sqrt{3}}{2} dy$, $e^3 = dy$, $e^4 = dz$. See Sec 5.7 of
  \refcite{excitations} for a 2-foliated X-cube model.}\BibitemShut {Stop}%
\bibitem [{Gfo()}]{Gfoot:XCubeEquiv}%
  \BibitemOpen
  \href@noop {} {}\bibinfo {note} {One can show \cite{Hfoot:mapping} that the
  FQFT with $M_k = N$ and $n_k=1$ for three flat foliations is dual to the
  X-cube QFT \cite{XcubeQFT,Seiberg3}. The FQFT is very closely related (see
  \appref{app:previous}{H}) to the foliated field theory in
  \refcite{stringMembraneNet}, in which an explicit connection to a
  string-membrane-net lattice model was shown in Sec. 3.3.3, and Sec. 3.3.2
  proved that the string-membrane-net model has the same ground states (up to
  generalized local unitary \cite{XieGLU}) as the X-cube model.}\BibitemShut
  {Stop}%
\bibitem [{\citenamefont {{Vijay}}(2017)}]{VijayLayer}%
  \BibitemOpen
  \bibfield  {author} {\bibinfo {author} {\bibfnamefont {S.}~\bibnamefont
  {{Vijay}}},\ }\href@noop {} {\  (\bibinfo {year} {2017})},\ \Eprint
  {http://arxiv.org/abs/1701.00762} {arXiv:1701.00762} \BibitemShut {NoStop}%
\bibitem [{\citenamefont {{Ma}}\ \emph {et~al.}(2017)\citenamefont {{Ma}},
  \citenamefont {{Lake}}, \citenamefont {{Chen}},\ and\ \citenamefont
  {{Hermele}}}]{MaLayer}%
  \BibitemOpen
  \bibfield  {author} {\bibinfo {author} {\bibfnamefont {H.}~\bibnamefont
  {{Ma}}}, \bibinfo {author} {\bibfnamefont {E.}~\bibnamefont {{Lake}}},
  \bibinfo {author} {\bibfnamefont {X.}~\bibnamefont {{Chen}}}, \ and\ \bibinfo
  {author} {\bibfnamefont {M.}~\bibnamefont {{Hermele}}},\ }\href {\doibase
  10.1103/PhysRevB.95.245126} {\bibfield  {journal} {\bibinfo  {journal}
  {\prb}\ }\textbf {\bibinfo {volume} {95}},\ \bibinfo {eid} {245126} (\bibinfo
  {year} {2017})},\ \Eprint {http://arxiv.org/abs/1701.00747}
  {arXiv:1701.00747} \BibitemShut {NoStop}%
\bibitem [{\citenamefont {{Prem}}\ \emph {et~al.}(2019)\citenamefont {{Prem}},
  \citenamefont {{Huang}}, \citenamefont {{Song}},\ and\ \citenamefont
  {{Hermele}}}]{CageNet}%
  \BibitemOpen
  \bibfield  {author} {\bibinfo {author} {\bibfnamefont {A.}~\bibnamefont
  {{Prem}}}, \bibinfo {author} {\bibfnamefont {S.-J.}\ \bibnamefont {{Huang}}},
  \bibinfo {author} {\bibfnamefont {H.}~\bibnamefont {{Song}}}, \ and\ \bibinfo
  {author} {\bibfnamefont {M.}~\bibnamefont {{Hermele}}},\ }\href {\doibase
  10.1103/PhysRevX.9.021010} {\bibfield  {journal} {\bibinfo  {journal}
  {Physical Review X}\ }\textbf {\bibinfo {volume} {9}},\ \bibinfo {eid}
  {021010} (\bibinfo {year} {2019})},\ \Eprint
  {http://arxiv.org/abs/1806.04687} {arXiv:1806.04687} \BibitemShut {NoStop}%
\bibitem [{Ffo()}]{Ffoot:appEoM}%
  \BibitemOpen
  \href@noop {} {}\bibinfo {note} {See \appref{app:quantized EoM}{D1} for
  details.}\BibitemShut {Stop}%
\bibitem [{\citenamefont {Hardorp}(1980)}]{totalFoliation}%
  \BibitemOpen
  \bibfield  {author} {\bibinfo {author} {\bibfnamefont {D.}~\bibnamefont
  {Hardorp}},\ }\href {\doibase 10.1090/memo/0233} {\emph {\bibinfo {title}
  {All compact orientable three dimensional manifolds admit total
  foliations}}},\ \bibinfo {series} {Memoirs of the American Mathematical
  Society}, Vol.~\bibinfo {volume} {26}\ (\bibinfo  {publisher} {American
  Mathematical Society},\ \bibinfo {year} {1980})\BibitemShut {NoStop}%
\bibitem [{Afo()}]{Afoot:twisted}%
  \BibitemOpen
  \href@noop {} {}\bibinfo {note} {See
  \refcite{twisted,gauging,majoranaCheckerboard,ShirleyCheckerboard} for more
  examples of foliated fracton orders.}\BibitemShut {Stop}%
\bibitem [{\citenamefont {{Williamson}}\ and\ \citenamefont
  {{Cheng}}(2020)}]{WilliamsonNonAbelianDefects}%
  \BibitemOpen
  \bibfield  {author} {\bibinfo {author} {\bibfnamefont {D.~J.}\ \bibnamefont
  {{Williamson}}}\ and\ \bibinfo {author} {\bibfnamefont {M.}~\bibnamefont
  {{Cheng}}},\ }\href@noop {} {\  (\bibinfo {year} {2020})},\ \Eprint
  {http://arxiv.org/abs/2004.07251} {arXiv:2004.07251} \BibitemShut {NoStop}%
\bibitem [{\citenamefont {{Prem}}\ and\ \citenamefont
  {{Williamson}}(2019)}]{PremPermutationNonAbelian}%
  \BibitemOpen
  \bibfield  {author} {\bibinfo {author} {\bibfnamefont {A.}~\bibnamefont
  {{Prem}}}\ and\ \bibinfo {author} {\bibfnamefont {D.}~\bibnamefont
  {{Williamson}}},\ }\href {\doibase 10.21468/SciPostPhys.7.5.068} {\bibfield
  {journal} {\bibinfo  {journal} {SciPost Physics}\ }\textbf {\bibinfo {volume}
  {7}},\ \bibinfo {eid} {068} (\bibinfo {year} {2019})},\ \Eprint
  {http://arxiv.org/abs/1905.06309} {arXiv:1905.06309} \BibitemShut {NoStop}%
\bibitem [{\citenamefont {{Bulmash}}\ and\ \citenamefont
  {{Barkeshli}}(2019)}]{BulmashNonAbelian}%
  \BibitemOpen
  \bibfield  {author} {\bibinfo {author} {\bibfnamefont {D.}~\bibnamefont
  {{Bulmash}}}\ and\ \bibinfo {author} {\bibfnamefont {M.}~\bibnamefont
  {{Barkeshli}}},\ }\href {\doibase 10.1103/PhysRevB.100.155146} {\bibfield
  {journal} {\bibinfo  {journal} {\prb}\ }\textbf {\bibinfo {volume} {100}},\
  \bibinfo {eid} {155146} (\bibinfo {year} {2019})},\ \Eprint
  {http://arxiv.org/abs/1905.05771} {arXiv:1905.05771} \BibitemShut {NoStop}%
\bibitem [{\citenamefont {Stephen}\ \emph {et~al.}(2020)\citenamefont
  {Stephen}, \citenamefont {Garre-Rubio}, \citenamefont {Dua},\ and\
  \citenamefont {Williamson}}]{DuaSSET}%
  \BibitemOpen
  \bibfield  {author} {\bibinfo {author} {\bibfnamefont {D.~T.}\ \bibnamefont
  {Stephen}}, \bibinfo {author} {\bibfnamefont {J.}~\bibnamefont
  {Garre-Rubio}}, \bibinfo {author} {\bibfnamefont {A.}~\bibnamefont {Dua}}, \
  and\ \bibinfo {author} {\bibfnamefont {D.~J.}\ \bibnamefont {Williamson}},\
  }\href {\doibase 10.1103/PhysRevResearch.2.033331} {\bibfield  {journal}
  {\bibinfo  {journal} {Phys. Rev. Research}\ }\textbf {\bibinfo {volume}
  {2}},\ \bibinfo {pages} {033331} (\bibinfo {year} {2020})},\ \Eprint
  {http://arxiv.org/abs/2004.04181} {arXiv:2004.04181} \BibitemShut {NoStop}%
\bibitem [{\citenamefont {{Tantivasadakarn}}\ and\ \citenamefont
  {{Vijay}}(2020)}]{TantivasadakarnSymmetryEnriched}%
  \BibitemOpen
  \bibfield  {author} {\bibinfo {author} {\bibfnamefont {N.}~\bibnamefont
  {{Tantivasadakarn}}}\ and\ \bibinfo {author} {\bibfnamefont {S.}~\bibnamefont
  {{Vijay}}},\ }\href {\doibase 10.1103/PhysRevB.101.165143} {\bibfield
  {journal} {\bibinfo  {journal} {\prb}\ }\textbf {\bibinfo {volume} {101}},\
  \bibinfo {eid} {165143} (\bibinfo {year} {2020})},\ \Eprint
  {http://arxiv.org/abs/1912.02826} {arXiv:1912.02826} \BibitemShut {NoStop}%
\bibitem [{\citenamefont {Devakul}\ \emph {et~al.}(2020)\citenamefont
  {Devakul}, \citenamefont {Shirley},\ and\ \citenamefont
  {Wang}}]{ShirleySSPTdual}%
  \BibitemOpen
  \bibfield  {author} {\bibinfo {author} {\bibfnamefont {T.}~\bibnamefont
  {Devakul}}, \bibinfo {author} {\bibfnamefont {W.}~\bibnamefont {Shirley}}, \
  and\ \bibinfo {author} {\bibfnamefont {J.}~\bibnamefont {Wang}},\ }\href
  {\doibase 10.1103/PhysRevResearch.2.012059} {\bibfield  {journal} {\bibinfo
  {journal} {Phys. Rev. Research}\ }\textbf {\bibinfo {volume} {2}},\ \bibinfo
  {pages} {012059} (\bibinfo {year} {2020})},\ \Eprint
  {http://arxiv.org/abs/1910.01630} {arXiv:1910.01630} \BibitemShut {NoStop}%
\bibitem [{\citenamefont {{Bulmash}}\ and\ \citenamefont
  {{Barkeshli}}(2018)}]{BulmashHiggs}%
  \BibitemOpen
  \bibfield  {author} {\bibinfo {author} {\bibfnamefont {D.}~\bibnamefont
  {{Bulmash}}}\ and\ \bibinfo {author} {\bibfnamefont {M.}~\bibnamefont
  {{Barkeshli}}},\ }\href {\doibase 10.1103/PhysRevB.97.235112} {\bibfield
  {journal} {\bibinfo  {journal} {\prb}\ }\textbf {\bibinfo {volume} {97}},\
  \bibinfo {eid} {235112} (\bibinfo {year} {2018})},\ \Eprint
  {http://arxiv.org/abs/1802.10099} {arXiv:1802.10099} \BibitemShut {NoStop}%
\bibitem [{\citenamefont {{Ma}}\ \emph {et~al.}(2018)\citenamefont {{Ma}},
  \citenamefont {{Hermele}},\ and\ \citenamefont {{Chen}}}]{MaHiggs}%
  \BibitemOpen
  \bibfield  {author} {\bibinfo {author} {\bibfnamefont {H.}~\bibnamefont
  {{Ma}}}, \bibinfo {author} {\bibfnamefont {M.}~\bibnamefont {{Hermele}}}, \
  and\ \bibinfo {author} {\bibfnamefont {X.}~\bibnamefont {{Chen}}},\ }\href
  {\doibase 10.1103/PhysRevB.98.035111} {\bibfield  {journal} {\bibinfo
  {journal} {\prb}\ }\textbf {\bibinfo {volume} {98}},\ \bibinfo {eid} {035111}
  (\bibinfo {year} {2018})},\ \Eprint {http://arxiv.org/abs/1802.10108}
  {arXiv:1802.10108} \BibitemShut {NoStop}%
\bibitem [{\citenamefont {{Radicevic}}(2019)}]{SystematicRadicevic}%
  \BibitemOpen
  \bibfield  {author} {\bibinfo {author} {\bibfnamefont {D.}~\bibnamefont
  {{Radicevic}}},\ }\href@noop {} {\  (\bibinfo {year} {2019})},\ \Eprint
  {http://arxiv.org/abs/1910.06336} {arXiv:1910.06336} \BibitemShut {NoStop}%
\bibitem [{\citenamefont {{Slagle}}\ \emph
  {et~al.}(2019{\natexlab{b}})\citenamefont {{Slagle}}, \citenamefont
  {{Prem}},\ and\ \citenamefont {{Pretko}}}]{SlagleCurvedU1}%
  \BibitemOpen
  \bibfield  {author} {\bibinfo {author} {\bibfnamefont {K.}~\bibnamefont
  {{Slagle}}}, \bibinfo {author} {\bibfnamefont {A.}~\bibnamefont {{Prem}}}, \
  and\ \bibinfo {author} {\bibfnamefont {M.}~\bibnamefont {{Pretko}}},\ }\href
  {\doibase 10.1016/j.aop.2019.167910} {\bibfield  {journal} {\bibinfo
  {journal} {Annals of Physics}\ }\textbf {\bibinfo {volume} {410}},\ \bibinfo
  {eid} {167910} (\bibinfo {year} {2019}{\natexlab{b}})},\ \Eprint
  {http://arxiv.org/abs/1807.00827} {arXiv:1807.00827} \BibitemShut {NoStop}%
\bibitem [{\citenamefont {{Bulmash}}\ and\ \citenamefont
  {{Iadecola}}(2019)}]{BulmashBoundary}%
  \BibitemOpen
  \bibfield  {author} {\bibinfo {author} {\bibfnamefont {D.}~\bibnamefont
  {{Bulmash}}}\ and\ \bibinfo {author} {\bibfnamefont {T.}~\bibnamefont
  {{Iadecola}}},\ }\href {\doibase 10.1103/PhysRevB.99.125132} {\bibfield
  {journal} {\bibinfo  {journal} {\prb}\ }\textbf {\bibinfo {volume} {99}},\
  \bibinfo {eid} {125132} (\bibinfo {year} {2019})},\ \Eprint
  {http://arxiv.org/abs/1810.00012} {arXiv:1810.00012} \BibitemShut {NoStop}%
\bibitem [{\citenamefont {{Li}}\ and\ \citenamefont
  {{Ye}}(2020)}]{PengYeDimensions}%
  \BibitemOpen
  \bibfield  {author} {\bibinfo {author} {\bibfnamefont {M.-Y.}\ \bibnamefont
  {{Li}}}\ and\ \bibinfo {author} {\bibfnamefont {P.}~\bibnamefont {{Ye}}},\
  }\href {\doibase 10.1103/PhysRevB.101.245134} {\bibfield  {journal} {\bibinfo
   {journal} {\prb}\ }\textbf {\bibinfo {volume} {101}},\ \bibinfo {eid}
  {245134} (\bibinfo {year} {2020})},\ \Eprint
  {http://arxiv.org/abs/1909.02814} {arXiv:1909.02814} \BibitemShut {NoStop}%
\bibitem [{\citenamefont {{Yoshida}}(2013)}]{YoshidaCode}%
  \BibitemOpen
  \bibfield  {author} {\bibinfo {author} {\bibfnamefont {B.}~\bibnamefont
  {{Yoshida}}},\ }\href {\doibase 10.1103/PhysRevB.88.125122} {\bibfield
  {journal} {\bibinfo  {journal} {\prb}\ }\textbf {\bibinfo {volume} {88}},\
  \bibinfo {eid} {125122} (\bibinfo {year} {2013})},\ \Eprint
  {http://arxiv.org/abs/1302.6248} {arXiv:1302.6248} \BibitemShut {NoStop}%
\bibitem [{\citenamefont {{Tian}}\ \emph {et~al.}(2020)\citenamefont {{Tian}},
  \citenamefont {{Samperton}},\ and\ \citenamefont {{Wang}}}]{ZhenghanHaah}%
  \BibitemOpen
  \bibfield  {author} {\bibinfo {author} {\bibfnamefont {K.~T.}\ \bibnamefont
  {{Tian}}}, \bibinfo {author} {\bibfnamefont {E.}~\bibnamefont {{Samperton}}},
  \ and\ \bibinfo {author} {\bibfnamefont {Z.}~\bibnamefont {{Wang}}},\ }\href
  {\doibase 10.1016/j.aop.2019.168014} {\bibfield  {journal} {\bibinfo
  {journal} {Annals of Physics}\ }\textbf {\bibinfo {volume} {412}},\ \bibinfo
  {eid} {168014} (\bibinfo {year} {2020})},\ \Eprint
  {http://arxiv.org/abs/1812.02101} {arXiv:1812.02101} \BibitemShut {NoStop}%
\bibitem [{\citenamefont {{Banks}}\ and\ \citenamefont
  {{Seiberg}}(2011)}]{SeibergBFZn}%
  \BibitemOpen
  \bibfield  {author} {\bibinfo {author} {\bibfnamefont {T.}~\bibnamefont
  {{Banks}}}\ and\ \bibinfo {author} {\bibfnamefont {N.}~\bibnamefont
  {{Seiberg}}},\ }\href {\doibase 10.1103/PhysRevD.83.084019} {\bibfield
  {journal} {\bibinfo  {journal} {\prd}\ }\textbf {\bibinfo {volume} {83}},\
  \bibinfo {eid} {084019} (\bibinfo {year} {2011})},\ \Eprint
  {http://arxiv.org/abs/1011.5120} {arXiv:1011.5120} \BibitemShut {NoStop}%
\bibitem [{\citenamefont {{Shirley}}\ \emph
  {et~al.}(2019{\natexlab{a}})\citenamefont {{Shirley}}, \citenamefont
  {{Slagle}},\ and\ \citenamefont {{Chen}}}]{excitations}%
  \BibitemOpen
  \bibfield  {author} {\bibinfo {author} {\bibfnamefont {W.}~\bibnamefont
  {{Shirley}}}, \bibinfo {author} {\bibfnamefont {K.}~\bibnamefont {{Slagle}}},
  \ and\ \bibinfo {author} {\bibfnamefont {X.}~\bibnamefont {{Chen}}},\ }\href
  {\doibase 10.1016/j.aop.2019.167922} {\bibfield  {journal} {\bibinfo
  {journal} {Annals of Physics}\ }\textbf {\bibinfo {volume} {410}},\ \bibinfo
  {eid} {167922} (\bibinfo {year} {2019}{\natexlab{a}})},\ \Eprint
  {http://arxiv.org/abs/1806.08625} {arXiv:1806.08625} \BibitemShut {NoStop}%
\bibitem [{\citenamefont {{Shao}}\ \emph {et~al.}()\citenamefont {{Shao}},
  \citenamefont {{Gorantla}}, \citenamefont {{Tat Lam}},\ and\ \citenamefont
  {{Seiberg}}}]{Hfoot:mapping}%
  \BibitemOpen
  \bibfield  {author} {\bibinfo {author} {\bibfnamefont {S.-H.}\ \bibnamefont
  {{Shao}}}, \bibinfo {author} {\bibfnamefont {P.}~\bibnamefont {{Gorantla}}},
  \bibinfo {author} {\bibfnamefont {H.}~\bibnamefont {{Tat Lam}}}, \ and\
  \bibinfo {author} {\bibfnamefont {N.}~\bibnamefont {{Seiberg}}},\ }\href@noop
  {} {}\bibinfo {note} {Private communication}\BibitemShut {NoStop}%
\bibitem [{\citenamefont {{Chen}}\ \emph {et~al.}(2010)\citenamefont {{Chen}},
  \citenamefont {{Gu}},\ and\ \citenamefont {{Wen}}}]{XieGLU}%
  \BibitemOpen
  \bibfield  {author} {\bibinfo {author} {\bibfnamefont {X.}~\bibnamefont
  {{Chen}}}, \bibinfo {author} {\bibfnamefont {Z.-C.}\ \bibnamefont {{Gu}}}, \
  and\ \bibinfo {author} {\bibfnamefont {X.-G.}\ \bibnamefont {{Wen}}},\ }\href
  {\doibase 10.1103/PhysRevB.82.155138} {\bibfield  {journal} {\bibinfo
  {journal} {\prb}\ }\textbf {\bibinfo {volume} {82}},\ \bibinfo {eid} {155138}
  (\bibinfo {year} {2010})},\ \Eprint {http://arxiv.org/abs/1004.3835}
  {arXiv:1004.3835} \BibitemShut {NoStop}%
\bibitem [{\citenamefont {{Wang}}\ \emph {et~al.}(2019)\citenamefont {{Wang}},
  \citenamefont {{Shirley}},\ and\ \citenamefont
  {{Chen}}}]{majoranaCheckerboard}%
  \BibitemOpen
  \bibfield  {author} {\bibinfo {author} {\bibfnamefont {T.}~\bibnamefont
  {{Wang}}}, \bibinfo {author} {\bibfnamefont {W.}~\bibnamefont {{Shirley}}}, \
  and\ \bibinfo {author} {\bibfnamefont {X.}~\bibnamefont {{Chen}}},\ }\href
  {\doibase 10.1103/PhysRevB.100.085127} {\bibfield  {journal} {\bibinfo
  {journal} {\prb}\ }\textbf {\bibinfo {volume} {100}},\ \bibinfo {eid}
  {085127} (\bibinfo {year} {2019})},\ \Eprint
  {http://arxiv.org/abs/1904.01111} {arXiv:1904.01111} \BibitemShut {NoStop}%
\bibitem [{\citenamefont {{Shirley}}\ \emph
  {et~al.}(2019{\natexlab{b}})\citenamefont {{Shirley}}, \citenamefont
  {{Slagle}},\ and\ \citenamefont {{Chen}}}]{ShirleyCheckerboard}%
  \BibitemOpen
  \bibfield  {author} {\bibinfo {author} {\bibfnamefont {W.}~\bibnamefont
  {{Shirley}}}, \bibinfo {author} {\bibfnamefont {K.}~\bibnamefont {{Slagle}}},
  \ and\ \bibinfo {author} {\bibfnamefont {X.}~\bibnamefont {{Chen}}},\ }\href
  {\doibase 10.1103/PhysRevB.99.115123} {\bibfield  {journal} {\bibinfo
  {journal} {\prb}\ }\textbf {\bibinfo {volume} {99}},\ \bibinfo {eid} {115123}
  (\bibinfo {year} {2019}{\natexlab{b}})},\ \Eprint
  {http://arxiv.org/abs/1806.08633} {arXiv:1806.08633} \BibitemShut {NoStop}%
\bibitem [{\citenamefont {{Witten}}(2008)}]{WittenGauge}%
  \BibitemOpen
  \bibfield  {author} {\bibinfo {author} {\bibfnamefont {E.}~\bibnamefont
  {{Witten}}},\ }\href@noop {} {\  (\bibinfo {year} {2008})},\ \Eprint
  {http://arxiv.org/abs/0812.4512} {arXiv:0812.4512} \BibitemShut {NoStop}%
\bibitem [{\citenamefont {Dijkgraaf}\ and\ \citenamefont
  {Witten}(1990)}]{DijkgraafWitten}%
  \BibitemOpen
  \bibfield  {author} {\bibinfo {author} {\bibfnamefont {R.}~\bibnamefont
  {Dijkgraaf}}\ and\ \bibinfo {author} {\bibfnamefont {E.}~\bibnamefont
  {Witten}},\ }\href {https://projecteuclid.org/euclid.cmp/1104180750}
  {\bibfield  {journal} {\bibinfo  {journal} {Comm. Math. Phys.}\ }\textbf
  {\bibinfo {volume} {129}},\ \bibinfo {pages} {393} (\bibinfo {year}
  {1990})}\BibitemShut {NoStop}%
\bibitem [{\citenamefont {{Kapustin}}\ and\ \citenamefont
  {{Seiberg}}(2014)}]{TQFTQFT}%
  \BibitemOpen
  \bibfield  {author} {\bibinfo {author} {\bibfnamefont {A.}~\bibnamefont
  {{Kapustin}}}\ and\ \bibinfo {author} {\bibfnamefont {N.}~\bibnamefont
  {{Seiberg}}},\ }\href {\doibase 10.1007/JHEP04(2014)001} {\bibfield
  {journal} {\bibinfo  {journal} {Journal of High Energy Physics}\ }\textbf
  {\bibinfo {volume} {2014}},\ \bibinfo {eid} {1} (\bibinfo {year} {2014})},\
  \Eprint {http://arxiv.org/abs/1401.0740} {arXiv:1401.0740} \BibitemShut
  {NoStop}%
\bibitem [{\citenamefont {Cordova}\ \emph {et~al.}(2020)\citenamefont
  {Cordova}, \citenamefont {Freed}, \citenamefont {Lam},\ and\ \citenamefont
  {Seiberg}}]{zeroForm}%
  \BibitemOpen
  \bibfield  {author} {\bibinfo {author} {\bibfnamefont {C.}~\bibnamefont
  {Cordova}}, \bibinfo {author} {\bibfnamefont {D.~S.}\ \bibnamefont {Freed}},
  \bibinfo {author} {\bibfnamefont {H.~T.}\ \bibnamefont {Lam}}, \ and\
  \bibinfo {author} {\bibfnamefont {N.}~\bibnamefont {Seiberg}},\ }\href
  {\doibase 10.21468/SciPostPhys.8.1.001} {\bibfield  {journal} {\bibinfo
  {journal} {SciPost Phys.}\ }\textbf {\bibinfo {volume} {8}},\ \bibinfo
  {pages} {1} (\bibinfo {year} {2020})},\ \Eprint
  {http://arxiv.org/abs/1905.09315} {arXiv:1905.09315} \BibitemShut {NoStop}%
\bibitem [{Bfo()}]{Bfoot:currents}%
  \BibitemOpen
  \href@noop {} {}\bibinfo {note} {Similar current constraints were given in
  Eqs. (7-8) of \refcite{stringMembraneNet}.}\BibitemShut {Stop}%
\bibitem [{\citenamefont {{Vidal}}(2007)}]{ERG}%
  \BibitemOpen
  \bibfield  {author} {\bibinfo {author} {\bibfnamefont {G.}~\bibnamefont
  {{Vidal}}},\ }\href {\doibase 10.1103/PhysRevLett.99.220405} {\bibfield
  {journal} {\bibinfo  {journal} {\prl}\ }\textbf {\bibinfo {volume} {99}},\
  \bibinfo {eid} {220405} (\bibinfo {year} {2007})},\ \Eprint
  {http://arxiv.org/abs/cond-mat/0512165} {arXiv:cond-mat/0512165} \BibitemShut
  {NoStop}%
\bibitem [{\citenamefont {{Aguado}}\ and\ \citenamefont
  {{Vidal}}(2008)}]{toricRG}%
  \BibitemOpen
  \bibfield  {author} {\bibinfo {author} {\bibfnamefont {M.}~\bibnamefont
  {{Aguado}}}\ and\ \bibinfo {author} {\bibfnamefont {G.}~\bibnamefont
  {{Vidal}}},\ }\href {\doibase 10.1103/PhysRevLett.100.070404} {\bibfield
  {journal} {\bibinfo  {journal} {\prl}\ }\textbf {\bibinfo {volume} {100}},\
  \bibinfo {eid} {070404} (\bibinfo {year} {2008})},\ \Eprint
  {http://arxiv.org/abs/0712.0348} {arXiv:0712.0348} \BibitemShut {NoStop}%
\bibitem [{\citenamefont {{Dua}}\ \emph {et~al.}(2020)\citenamefont {{Dua}},
  \citenamefont {{Sarkar}}, \citenamefont {{Williamson}},\ and\ \citenamefont
  {{Cheng}}}]{DuaBifurcating}%
  \BibitemOpen
  \bibfield  {author} {\bibinfo {author} {\bibfnamefont {A.}~\bibnamefont
  {{Dua}}}, \bibinfo {author} {\bibfnamefont {P.}~\bibnamefont {{Sarkar}}},
  \bibinfo {author} {\bibfnamefont {D.~J.}\ \bibnamefont {{Williamson}}}, \
  and\ \bibinfo {author} {\bibfnamefont {M.}~\bibnamefont {{Cheng}}},\ }\href
  {\doibase 10.1103/PhysRevResearch.2.033021} {\bibfield  {journal} {\bibinfo
  {journal} {Physical Review Research}\ }\textbf {\bibinfo {volume} {2}},\
  \bibinfo {eid} {033021} (\bibinfo {year} {2020})},\ \Eprint
  {http://arxiv.org/abs/1909.12304} {arXiv:1909.12304} \BibitemShut {NoStop}%
\end{thebibliography}%

\arxiv{
\newpage
\appendix

\section{Gauge Fields (Review)}
\label{app:gauge field}

Here, we review the definition of a 1-form gauge field.
Mathematically, a 1-form gauge field with gauge group $G = U(1)$ is a connection on $\M \times G$
  (or more generally a $G$-bundle $E \to \M$),
  where $\M$ is the spacetime manifold. \cite{WittenGauge,DijkgraafWitten}.

This can be made more explicit by considering a good open cover of the spacetime manifold $\M$;
  i.e. consider a collection of sets $U_i \subset \M$ that cover $\M$
  (i.e. $\cup_i U_i = \M$)
  such that finite intersections $U_{i_1} \cap U_{i_2} \cap \cdots \cap U_{i_n}$ are diffeomorphic to an open ball.
A 1-form gauge field can then be specified by the following data and constraints \cite{TQFTQFT}:
(1) The gauge field is locally defined on each $U_i$ by a 1-form $A_{(i)}$.\footnote{%
  The parenthesis in $A_{(i)}$ are used to emphasize that $i$ is an index for a spacetime patch $U_i$;
    $i$ is not a coordinate index $\mu=0,1,2,3$.}
(2) On nonempty overlaps $U_i \cap U_j$ (depicted in yellow below), the two locally defined fields $A_{(i)}$ and $A_{(j)}$ must be equal up to a gauge transformation:
\begin{align}
  \vcenter{\hbox{\includegraphics[width=.15\columnwidth]{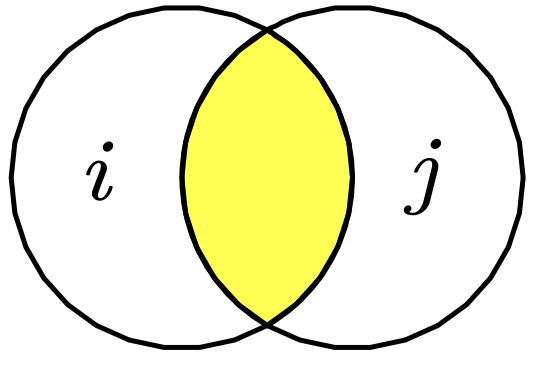}}} &&
  A_{(i)} - A_{(j)} = dg_{(ij)} && \label{eq:g}
\end{align}
where $g_{(ij)}$ are called transition functions.
(3) On nonempty triple-overlaps $U_i \cap U_j \cap U_k$ (depicted in yellow below), the transition functions must satisfy the cocycle condition up to an integer multiple of $2\pi$:
\begin{align}
  \vcenter{\hbox{\includegraphics[width=.2\columnwidth]{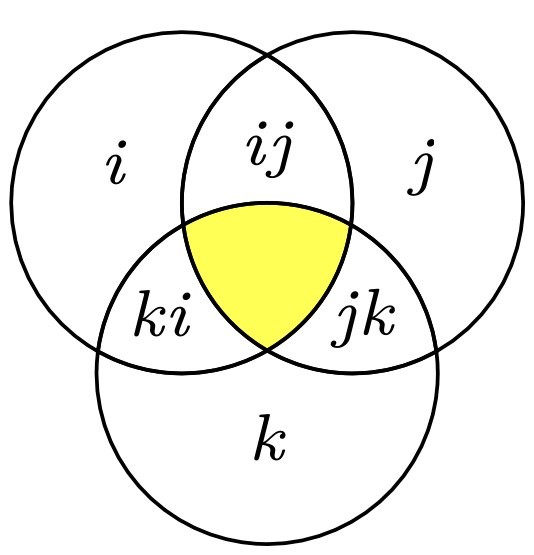}}} &&
  g_{(ij)} + g_{(jk)} + g_{(ki)} \in 2\pi\ZZ \label{eq:g sum}
\end{align}

A 0-form gauge field $\theta$ can be similarly defined by
  a 0-form $\theta_{(i)} : U_i \to \RR$ on each $U_i$ where
  $\theta_{(i)} - \theta_{(j)} \in 2\pi\ZZ$ on overlaps $U_i \cap U_j$.
Thus, $\theta$ could alternatively be defined as a $U(1)$-valued function $\tilde{\theta} : \M \to U(1)$.

See also the beginning of \refcite{TQFTQFT} for another a review of $q$-form gauge fields
  and Section 2.1 of \refcite{zeroForm} for an explicit example of how to define integrals of gauge fields.

%
% ...

\subsection{Example}
\label{eq:gauge field example}

Here, we review a simple example of a field configuration for a trivial $2\pi$ flux.
Consider BF theory on a 2+1D torus:
  $L = \frac{N}{2\pi} B \wedge dA$ where $A$ and $B$ are 1-form gauge fields.
Decompose the 3-torus as $\M_\trm \times \M_\xrm \times \M_\yrm$ with lengths $\lt$, $\lx$, and $\ly$,
  where $\M_\xrm$ is a circle with coordinate $x \in [0,\lx)$,
  and similar for $\M_\trm$ and $\M_\yrm$.
Then a $2\pi$ flux that is evenly spread throughout space will have
\begin{equation}
  dA=\frac{2\pi}{\lx \ly} \dd x \wedge \dd y
\end{equation}

To formally specify the gauge field $A$, first choose an open cover\footnote{%
    \eqnref{eq:3cover} is not a \emph{good} open cover [defined above \eqnref{eq:g}]
      since e.g. $U_i$ and $U_i \cap U_j$ are not diffeomorphic to an open ball.
    However, this open cover is sufficient for this example,
      and it is trivial (but tedious)
      to extend this open cover to a good open cover by shrinking and adding more submanifolds $U_i$.}
  given by (\figref{fig:cover}):
\begin{equation}
\begin{array}{ccccccc}
  U_1 & = & \M_\trm & \times & (0,\lx) & \times & (0,\ly) \\
  U_2 & = & \M_\trm & \times & [0,\frac{\lx}{2}) \cup (\frac{\lx}{2},\lx) & \times & (0,\ly) \\
  U_3 & = & \M_\trm & \times & (0,\lx) & \times & [0,\frac{\ly}{2}) \cup (\frac{\ly}{2},\ly) \\
  U_4 & = & \M_\trm & \times & [0,\frac{\lx}{2}) \cup (\frac{\lx}{2},\lx) & \times & [0,\frac{\ly}{2}) \cup (\frac{\ly}{2},\ly)
\end{array} \label{eq:3cover}
\end{equation}
Note that $0 \in [0,\lx)$ while $0 \notin (0,\lx)$.
Now the gauge field can be defined by
\begin{equation}
\begin{aligned}
  A_{(1)} &= A_{(3)} = \frac{2\pi}{\lx \ly} x \, \dd y \\
  A_{(2)} &= A_{(4)} = \frac{2\pi}{\lx \ly}
              \begin{cases}  x      \, \dd y & 0 \leq x < \frac{\lx}{2} \\
                            (x-\lx) \, \dd y & \frac{\lx}{2} < x \leq \lx \end{cases}
\end{aligned}
\end{equation}
with transition functions
\begin{align}
  g_{(12)} &= g_{(14)} = g_{(32)} = \begin{cases}
              0                  & 0 < x < \frac{\lx}{2} \\
              \frac{2\pi}{\ly} y & \frac{\lx}{2} < x < \lx \end{cases} \\
  g_{(34)} &= \begin{cases}
              0                  & 0 < x < \frac{\lx}{2} \\
              \frac{2\pi}{\ly} y & \frac{\lx}{2} < x < \lx \text{ and } 0 \leq y < \frac{\ly}{2} \\
              \frac{2\pi}{\ly} (y-\ly) & \frac{\lx}{2} < x < \lx \text{ and } \frac{\ly}{2} < y \leq \ly \end{cases} \nn\\
  g_{(13)} &= g_{(24)} = 0 \nn
\end{align}
Note that $A_{(i)}$ and $g_{(ij)}$ are continuous and satisfy \eqnref{eq:g} and \eqref{eq:g sum}.
Also note that if we rescale $A_{(i)}$ and $g_{(ij)}$ by some constant $\alpha \in \RR$,
  then \eqnref{eq:g sum} will only be satisfied if $\alpha \in \ZZ$.
Therefore, the total flux $\int dA$ must be an integer multiple of $2\pi$, which is physically trivial.

\begin{figure}
  \includegraphics[width=.6\columnwidth]{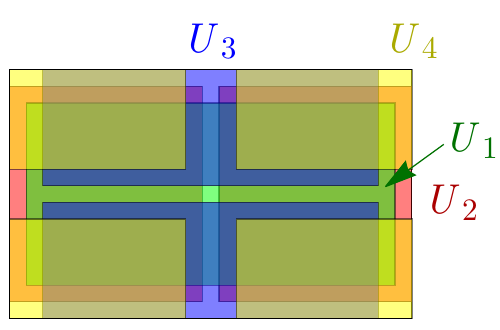}
  \caption{%
    An XY planar slice of spacetime showing a depiction of $U_1$, $U_2$, $U_3$, and $U_4$
      from \eqsref{eq:3cover} and \eqref{eq:3cover'}.
  }\label{fig:cover}
\end{figure}

\section{Foliated Gauge Fields}
\label{app:foliated gauge field}

Here, we provide a more formal definition of foliated gauge fields.
See \appref{app:gauge field}{} for a review of ordinary gauge fields.

We will provide two definitions, which we believe are equivalent.
The first definition is that a foliated gauge field $A$ is given by an ordinary gauge field $A_\ell$ on each leaf $\ell$ of a foliation.
Then the integral of a foliated $(q+1)$-form gauge field $A$ over a foliated $(q+1)$-dimensional manifold $\M$ is given by
  the infinite sum of integrals over each leaf $\ell \subset \M$ of the foliation:
  $\int_\M A \equiv \sum_\ell \int_\ell A_\ell$.

We now provide a second definition, which avoids the infinite summation over leaves.
This definition is also simpler locally (as it reduces to just a constrained 2-form gauge field).
We use this second definition throughout the rest of this text.
Consider a good open cover of sets $U_i \subset \M$ [as defined above \eqnref{eq:g}]
  that cover the spacetime manifold $\M$,
  which is foliated using a foliation field $e$\arxiv{, as defined in \secref{sec:foliation field}}.
A foliated (1+1)-form gauge field is defined by the following data and constraints:
(1) The foliated gauge field is locally defined on each $U_i$ by a 2-form field $A_{(i)}$ that obeys the constraint $A_{(i)} \wedge e = 0$
  [as in \eqnref{eq:A constraint}].
(2) On nonempty overlaps $U_i \cap U_j$, the two locally defined fields $A_{(i)}$ and $A_{(j)}$ must be equal up to a gauge transformation:
\begin{equation}
  A_{(i)} - A_{(j)} = dg_{(ij)} \label{eq:g'}
\end{equation}
where $g_{(ij)}$ is a foliated (0+1)-form transition function that obeys $g_{(ij)} \wedge e = 0$.
(3) On nonempty triple-overlaps $U_i \cap U_j \cap U_k$,
  these transition functions must satisfy a foliated cocycle condition:
\begin{align}
  \vcenter{\hbox{\includegraphics[width=.2\columnwidth]{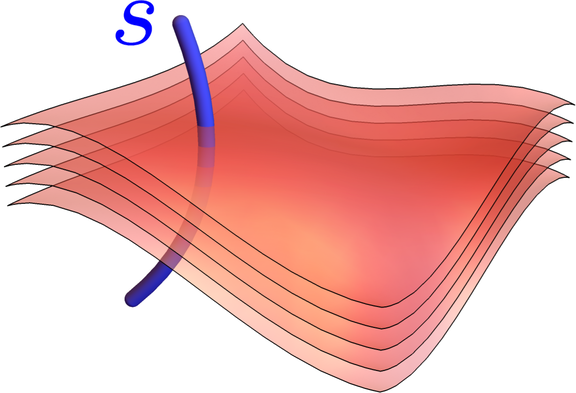}}} &&
  \int_s g_{(ij)} + g_{(jk)} + g_{(ki)} \in 2\pi\ZZ \label{eq:g sum'}
\end{align}
where $s$ is any 1D manifold (possibly with boundaries) transverse\footnote{%
  A 1-dimensional manifold is transverse to a foliation if it is never tangent to a leaf.
  } to the foliation.
An example is depicted in the graphic, with $s$ drawn as a blue line.

\subsection{Example}
\label{eq:foliated gauge field example}

Here, we demonstrate a foliated analog of the example in \appref{eq:gauge field example}{}.
That is, we wish to describe a field configuration for a trivial $2\pi$ flux on a single leaf of a foliation.
We will consider the following FQFT on a 3+1D torus:
\begin{equation}
  L = \frac{N}{2\pi} B \wedge dA
\end{equation}
where $A$ is a foliated (1+1)-form gauge field and $B$ is a 1-form gauge field.
This FQFT describes a foliation of 2+1D BF theories.

For the first definition, a $2\pi$ flux that is evenly spread throughout a leaf $\ell_0$ of the foliation will have:
\begin{equation}
  dA_\ell = \begin{cases}
              \frac{2\pi}{\lx \ly} \dd x \wedge \dd y & \ell = \ell_0 \\
               0                                      & \ell \neq \ell_0
            \end{cases}
\end{equation}
where $\ell$ indexes the different leaves of the foliation.
$A_{\ell_0}$ can then be defined as in \appref{eq:gauge field example}{}.

Now consider the second foliated gauge field definition.
For simplicity, consider a flat foliation with $e = \dd z$.
Decompose the 4-torus as $\M_\trm \times \M_\xrm \times \M_\yrm \times \M_\zrm$ with lengths $\lt$, $\lx$, $\ly$, and $\lz$,
  where $\M_\xrm$ is a circle with coordinate $x \in [0,\lx)$,
  and similar for $\M_\trm$, $\M_\yrm$, and $\M_\zrm$.
A $2\pi$ flux that is evenly spread throughout a leaf (at $z = z_0$) of the foliation will have:
\begin{equation}
  dA = \frac{2\pi}{\lx \ly} \delta(z-z_0) \, \dd x \wedge \dd y \wedge \dd z
\end{equation}
\begin{widetext}
To formally specify the foliated gauge field $A$, first choose an open cover given by (\figref{fig:cover}):
\begin{equation}
\begin{array}{ccccccccc}
  U_1 & = & \M_\trm & \times & (0,\lx) & \times & (0,\ly) & \times & \M_\zrm \\
  U_2 & = & \M_\trm & \times & [0,\frac{\lx}{2}) \cup (\frac{\lx}{2},\lx) & \times & (0,\ly) & \times & \M_\zrm \\
  U_3 & = & \M_\trm & \times & (0,\lx) & \times & [0,\frac{\ly}{2}) \cup (\frac{\ly}{2},\ly) & \times & \M_\zrm \\
  U_4 & = & \M_\trm & \times & [0,\frac{\lx}{2}) \cup (\frac{\lx}{2},\lx) & \times & [0,\frac{\ly}{2}) \cup (\frac{\ly}{2},\ly) & \times & \M_\zrm
\end{array} \label{eq:3cover'}
\end{equation}
Now the foliated gauge field can be defined by
\begin{equation}
\begin{aligned}
  A_{(1)} &= A_{(3)} = \frac{2\pi}{\lx \ly} \delta(z-z_0) x \, \dd y \wedge \dd z \\
  A_{(2)} &= A_{(4)} = \frac{2\pi}{\lx \ly} \delta(z-z_0)
              \begin{cases}  x      \, \dd y \wedge \dd z & 0 \leq x < \frac{\lx}{2} \\
                            (x-\lx) \, \dd y \wedge \dd z & \frac{\lx}{2} < x \leq \lx \end{cases}
\end{aligned}
\end{equation}
with transition functions
\begin{align}
  g_{(12)} &= g_{(14)} = g_{(32)} = \begin{cases}
                0 & 0 < x < \frac{\lx}{2} \\
                \frac{2\pi}{\ly} \delta(z-z_0) \, y \, \dd z
                  & \frac{\lx}{2} < x < \lx \end{cases} \nn\\
  g_{(34)} &= \begin{cases}
              0                  & 0 < x < \frac{\lx}{2} \\
              \frac{2\pi}{\ly} \delta(z-z_0) \, y \, \dd z & \frac{\lx}{2} < x < \lx \text{ and } 0 \leq y < \frac{\ly}{2} \\
              \frac{2\pi}{\ly} \delta(z-z_0) \, (y-\ly) \, \dd z & \frac{\lx}{2} < x < \lx \text{ and } \frac{\ly}{2} < y \leq \ly \end{cases} \\
  g_{(13)} &= g_{(24)} = 0 \nn
\end{align}
Note that $A_{(i)}$ and $g_{(ij)}$ satisfy \eqnref{eq:g'} and \eqref{eq:g sum'}.
\end{widetext}

\section{Mobility Constraints and Currents}
\label{app:mobility}

\arxivPR{In \secref{sec:operators}, we}{We} studied the rigidity of the gauge invariant operators.
This rigidity is analogous to the particle mobility constrains characteristic of fracton models.
Consider a more general operator of the form $e^{\ii \int L'}$ where:
\begin{equation}
  L' = - \sum_k A^k \wedge J^k - \sum_k B^k \wedge I^k - a \wedge j - b \wedge i \label{eq:L'}
\end{equation}
$J^k$ and $i$ are 2-forms;
  $I^k$ is a (2+1)-form (i.e. $I^k \wedge e^k = 0$);
  and $j$ is a 3-form.
$J^k$, $I^k$, $j$, and $i$ can be thought of as current sources that parameterize the generic operator $e^{\ii \int L'}$.

$e^{\ii \int L'}$ is only gauge invariant if the following mobility constraints are satisfied:
\begin{align}
  dJ^k \wedge e^k &= -m_k \, j \wedge e^k \label{eq:J mobility}\\
  dI^k &= 0,\quad I^k\wedge e^k=0 \label{eq:I mobility}\\
  dj&=0 \label{eq:j mobility}\\
  di&=\sum_k n_k I^k \label{eq:i mobility}
\end{align}
These constraints result from imposing gauge invariance under the
  $\zeta^k$, $\chi^k$, $\alpha^k$, $\lambda$, and $\mu$ transformations in \eqnref{eq:gauge}, respectively.
The local foliation field constraint $A^k \wedge e^k = 0$ [\eqnref{eq:A constraint}]
  also results in the following redundancy:
  $J^k \to J^k + \phi^k\wedge e^k$,
  where $\phi^k$ is an arbitrary 1-form.
This gives $J^k$ the same number of degrees of freedom as a (2+1)-form. \cite{Bfoot:currents}

When the FQFT describes $Z_N$ X-cube (i.e. when $n_k=1$ and $M_k=N$ with three foliations):
  $j$ is the fracton current,
  $J$ is a fracton dipole current,
  linear combinations of $I$ currents result in lineons,
  and $i$ is a current for string excitations which do not appear in the X-cube model\footnote{%
    However, it is possible to map the $i$ current to a string of many lineons using the mappings in Sections 3.3.2 and 3.3.3 of \refcite{stringMembraneNet}.}.

\eqnref{eq:J mobility} tells us that any $j$ current that passes through a leaf of a foliation must be compensated by the divergence of $J$ current.
This is analogous to the X-cube model where moving a fracton ($j$ current)
  requires creating fracton dipoles ($J$ divergence).

\eqnref{eq:i mobility} implies that the $i$ current describes string excitations.
If there are no string excitations (i.e. if $di=0$),
  then \eqnref{eq:i mobility} implies that $\sum_k I^k = 0$.
This implies that $I$ current must come in pairs.
$I^k\wedge e^k=0$ [\eqnref{eq:I mobility}] implies that the $I$ current can only move along a leaf of the $k^\text{th}$ foliation.
But since $I$ current must come in pairs for different foliations $k$,
  a particle of $I$ current must be bound to two leaves for two different foliations,
  which implies that $I$ describes currents of lineons.

\section{Equations of Motion}

The equations of motion for the FQFT Lagrangian coupled to source currents,
  $L+L'$ [\eqsref{eq:L} and \eqref{eq:L'}], are given below:
\begin{align}
  \frac{M_k}{2\pi}(dB^k + n_k b)\wedge e^k &= J^k \wedge e^k \label{eq:EoM J}\\
  \frac{M_k}{2\pi} dA^k &= I^k \label{eq:EoM I}\\
  \frac{N}{2\pi}db &= j \label{eq:EoM j}\\
  \frac{N}{2\pi}(da+\sum_k m_k A^k) &= i \label{eq:EoM i}
\end{align}

\subsection{Quantized Integrals}
\label{app:quantized EoM}

Since the gauge fields are compact,
  there are also nonlocal ``equations of motion'' that result in quantized integrals.

For example, let us derive the following quantized period from the main text [\eqnref{eq:quantized EoM}]:
\begin{equation}
  \oint_{\M_2} b \in \frac{2\pi}{N} \ZZ \label{eq:M2 EoM}
\end{equation}
where $\M_2$ is a closed 2-manifold.
If $\M_2$ is contractible, then $\oint_{\M_2} b = 0$ by the local equation of motion
  $db=0$ \eqref{eq:EoM j}.
Consider a simple example of non-contractible $\M_2$ on a
  spacetime manifold that is an $\lt \times \lx \times \ly \times \lz$ 4-torus.
Let $\M_2$ be a tz-plane.
Now consider summing over field configurations with flux
\begin{equation}
 f_Q = da = Q \frac{2\pi}{\lx\ly} \, \dd x \wedge \dd y \label{eq:fq}
\end{equation}
for all $Q \in \ZZ$
  (similar to the example in \appref{eq:gauge field example}{}).
Summing over this subset of field configurations shows that the partition function is zero unless the following integral is quantized:
  $\int \frac{N}{2\pi} b \wedge da \in 2\pi \ZZ$.
Therefore, $\int \frac{N}{2\pi} b \wedge da
  = \int \frac{N}{2\pi} b \wedge f_Q
  = Q \frac{N}{\lx\ly} \int_{x,y} \int_{t,z} b_{03}
  = Q N \int_{\M_2} b \in 2\pi\ZZ$,
  where the last equality follows because the integral of $b$ over any tz-plane will be equal
  due to the equation of motion $db=0$ \eqref{eq:EoM j}.
This demonstrates the quantization \eqref{eq:M2 EoM}.

We now derive the other quantized period in \eqnref{eq:quantized EoM}:
\begin{equation}
  \oint_{\M_2^\text{P}} A^k \in \frac{2\pi}{M_k}\ZZ \label{eq:M2P EoM}
\end{equation}
Consider the simple but nontrivial example where $\M_2^\text{P}$ is a tz-plane of a spacetime 4-torus,
  and suppose that the first foliation field is $e^1 = \dd z$.
Then similar to \eqnref{eq:fq},
  we can sum over fluxes $dB^1 = Q \frac{2\pi}{\lx\ly} \, \dd x \wedge \dd y$ for all $Q \in \ZZ$
  and apply $dA^1=0$ [\eqnref{eq:EoM I}]
  to derive \eqnref{eq:M2P EoM} with $k=1$.

We now derive the following quantized period\footnote{\label{foot:removed}%
  We will only demonstrate quantization of \eqnref{eq:M1F EoM} and \eqref{eq:M1L EoM}
    when $\M_1^\text{F}$ and $\M_1^\text{L}$ are removed from the respective spacetimes.
  This means that the gauge fields will not have to be well-defined or continuous on $\M_1^\text{F}$ or $\M_1^\text{L}$.
  This is sufficient for demonstrating the operator quantization in \arxiv{\secref{sec:operators} of} the main text.}:
\begin{equation}
  \oint_{\M_1^\text{F}} a \in \frac{2\pi}{N} \ZZ \label{eq:M1F EoM}
\end{equation}
where $\M_1^\text{F}$ is supported on a single leaf for each foliation with $n_k \neq 0$ [as in \eqnref{eq:fracton}].
Consider the simple but nontrivial example where $\M_1^\text{F}$ is a loop around a periodic time direction
  and centered at the origin of the spatial manifold $\RR^3$.
Then \eqnref{eq:M1F EoM} will result from summing over fluxes
  $db = Q \, 2\pi \, \delta^3({\bf x}) \, \dd x \wedge \dd y \wedge \dd z$ for all $Q \in \ZZ$
  and choosing $B^k$ such that $dB^k + n_k b = 0$.
These fluxes can be realized by
  $b = - Q \, \frac{1}{2} d(\cos\theta) \wedge \dd\phi$ and
  $B^k = n_k Q \, \frac{1}{2} (\cos\theta-1) \, \dd\phi$ in spherical coordinates.
Summing over this subset of field configurations shows that the partition function is zero
  unless $\int L \in 2\pi \ZZ$, where
  $\int L = - \int \frac{N}{2\pi} db \wedge a = Q N \int_{\M_1^\text{F}} a$.
This demonstrates the quantization \eqnref{eq:M1F EoM}.

Finally, we derive the following quantized period\arxiv{$^{\ref{foot:removed}}$}:
\begin{equation}
  \oint_{\M_1^\text{L}} \sum_k q_k B^k \in 2\pi \ZZ \text{ when } q_k \in M_k \ZZ \label{eq:M1L EoM}
\end{equation}
where $\sum_k q_k n_k = 0$ and $\M_1^\text{L}$ is supported on a single leaf for each foliation with
  $q_k \neq 0$ [as in \eqnref{eq:lineon}].
Consider the simple but nontrivial example where $\M_1^\text{L}$ is a loop around a periodic time direction
  and centered at the origin of the spatial manifold $\RR^3$.
Then \eqnref{eq:M1L EoM} will result from summing over fluxes
  $dA^k = \frac{q_k}{M_k} F_Q$ where $F_Q = Q \, 2\pi \, \delta^3({\bf x}) \, \dd x \wedge \dd y \wedge \dd z$
  for each $Q \in \ZZ$ with $a_\mu$ chosen such that $da + \sum_k m_k A^k = 0$.
To realize the flux $dA^k$, consider the example foliation $e^1 = \dd z$;
  then $A^1 = Q \frac{q_1}{M_1} \delta(z) \, \dd\phi \wedge \dd z$
  in polar coordinates (where $q_k \in M_k \ZZ$).
Then $d \left( \sum_k m_k A^k \right) = \sum_k m_k \frac{q_k}{M_k} F_Q
  = \frac{1}{N} F_Q \sum_k q_k n_k = 0$;
  therefore, it is possible to choose $a_\mu$ such that $da + \sum_k m_k A^k = 0$.
Summing over this subset of field configurations shows that the partition function is zero unless the following integral is quantized:
  $\int L = \int \sum_k \frac{M_k}{2\pi} B^k \wedge d A^k + \frac{N}{2\pi} b \wedge \left( d a + \sum_k m_k A^k \right)
    = \int \sum_k q_k B^k \wedge \frac{1}{2\pi} F_Q = Q \int_{\M_1^\text{L}} \sum_k q_k B^k \in 2\pi \ZZ$.

\section{Lineon Operator}
\label{app:lineon}

Consider the string operator $T$ in \eqnref{eq:lineon} with a charge vector $q_k$ such that $\sum_k q_k n_k = 0$.
Below, we prove that this charge vector can always be decomposed into a sum of lineon and planon charge vectors
  (which have at most two nonzero elements).

To prove this, first extract all planon charges\footnote{%
  $\delta_{k,k'} = 1$ if $k=k'$ else $\delta_{k,k'} = 0$.}
  $q^{(k')}_k \equiv q_k \delta_{k,k'}$
  from the charge vector $q_k$:
\begin{equation}
  q_k = q'_k + \sum_{k' \in K_\text{P}} q^{(k')}_k
\end{equation}
$\sum_{k' \in K_\text{P}}$ sums over all foliations $k'$ such that $q_{k'} \neq 0$ and $n_{k'} = 0$.
We are left with a new charge vector $q'_k$, such that
  $q'_k = 0$ for all $k$ with $n_k = 0$.
Next, we show that $q'_k$ can be decomposed into lineons.

If $q'_k$ has at most two nonzero components,
  then $q'_k$ is a lineon and the proof is complete.
Otherwise, without loss of generality (by reordering the foliations $k$),
  assume that $q'_1 \neq 0$.
Next, we show that $q'_k$ can be decomposed into $q''_k$ and lineon charge vectors $q^{(1,k')}_k$:
\begin{equation}
  q'_k = q''_k + \sum_{k' \in K_\text{L}} q^{(1,k')}_k \label{eq:q''}
\end{equation}
$\sum_{k' \in K_\text{L}}$ only sums over foliations $k'=2,3,\cdots,\nf$ such that $q'_{k'} \neq 0$.
$q^{(1,k')}_k \neq 0$ only for $k=1$ and $k=k'$.
Also note that all charge vectors in this proof ($q_k$, $q'_k$, $q''_k$, $q^{(k')}_k$, and $q^{(1,k')}_k$)
  are valid charge vectors (i.e. $\sum_k q'_k n_k = 0$, and similar for the other charge vectors).
We want to choose the $q^{(1,k')}_k$ such that
  $q''_1 = 0$ and $q''_k = 0$ for each $k$ with $q'_k = 0$.
Then $q''_k$ will have at least one more zero element than $q'_k$.
Thus, we can complete the proof by repeatedly reapplying the logic of this paragraph (with $q''_k \to q'_k$)
  until $q'_k$ is a lineon with two nonzero components.

We now just need to show that the decomposition in \eqnref{eq:q''} is possible.
Without loss of generality, assume $n_k \neq 0$ and $q'_k \neq 0$ for all foliations $k$
  (by just ignoring foliations $k$ for which this is not true).
Let
\begin{align}
  q^{(1,k')}_k &= Q^{(k')} \begin{cases}
    + r_{k',1} & k=1  \\
    - r_{1,k'} & k=k' \\
    0             & \text{otherwise} \end{cases} \\
  r_{k,k'} &\equiv \frac{n_k}{\gcd(n_k,n_{k'})}
\end{align}
for some integers $Q^{(k')}$.
$\gcd$ denotes the greatest common divisor.
We want $\sum_{k'=2}^{\nf} q^{(1,k')}_1 = q'_1$ so that $q''_1 = 0$ in \eqnref{eq:q''}.
By appropriately choosing $Q^{(k')} \in \ZZ$,
  the sum $\sum_{k'=2}^{\nf} q^{(1,k')}_1$ can be any integer multiple of $R_1 \equiv \gcd(r_{2,1}, r_{3,1}, \cdots r_{\nf,1})$.
Therefore, we just need to show that $q'_1$ is an integer multiple of $R_1$.
But $q'_1 n_1 = u R$ where $R \equiv \gcd(n_2, n_3, \cdots n_\nf)$ for some $u \in \ZZ$ since $\sum_k q_k n_k = 0$.
Thus, $q'_1 = \frac{u R}{n_1} = [u \gcd(R, n_1)/n_1] \, [R / \gcd(R, n_1)]$
  is an integer multiple of $R / \gcd(R, n_1)$.
But $R / \gcd(R, n_1) = R_1$ (by properties of $\gcd$ and integer division).
Therefore, $q'_1$ is an integer multiple of $R_1$, which completes the proof.


\section{Entanglement RG}
\label{app:RG}

Entanglement RG \cite{ERG} studies coarse graining by using a local unitary transformation to decouple degrees of freedom from a ground state.

For example, a local unitary can be used to coarse-grain the ground state (GS) of toric code on a periodic $2L \times 2L$ lattice to
  the toric code GS on a periodic $L \times L$ lattice along with decoupled qubits \cite{toricRG}:
\begin{gather}
  U \; \ket{2L \times 2L \text{ toric code GS}} \nn\\ = \\
  \ket{L\times L \text{ toric code GS}} \otimes \ket{\underbrace{\uparrow \cdots \uparrow}_{\times 6L^2}} \nn
\end{gather}

In a 3D foliated fracton order, the local unitary decouples 2D topological orders in addition to decoupled qubits.
For example, slightly coarse-graining the X-cube model in one direction exfoliates a decoupled layer of toric code \cite{3manifolds,DuaBifurcating}:
\begin{gather}
  U \; \ket{L_x \times L_y \times L_z \text{ X-cube GS}} \nn\\ = \label{eq:Xcube RG} \\
  \ket{L_x \times L_y \times (L_z-1) \text{ X-cube GS}} \otimes \nn\\
    \ket{L_x \times L_y \text{ toric code GS}} \otimes \ket{\underbrace{\uparrow \cdots \uparrow}_{\times L_x L_y}} \nn
\end{gather}
A local unitary can also be used to exfoliate every other layer:
\begin{gather}
  U \; \ket{L_x \times L_y \times L_z \text{ X-cube GS}} \nn\\ = \\
  \ket{L_x \times L_y \times \frac{L_z}{2} \text{ X-cube GS}} \nn\\
    \otimes_{k=1}^{L_z/2} \ket{L_x \times L_y \text{ toric code GS}} \otimes \ket{\underbrace{\uparrow \cdots \uparrow}_{\times L_x L_y \frac{L_z}{2}}} \nn
\end{gather}

Entanglement RG is convenient for exactly solvable lattice models since the RG can often be done exactly using a simple formalism.
Entanglement RG is also useful because it only discards degrees of freedom after they have been explicitly decoupled.
This is in contrast to Wilsonian RG,
  where one could (in principal) accidentally integrate out important degrees of freedom.

\section{Exfoliation}
\label{app:exfoliation}

\begin{figure}
  \subfloat[]{\includegraphics[width=.35\columnwidth]{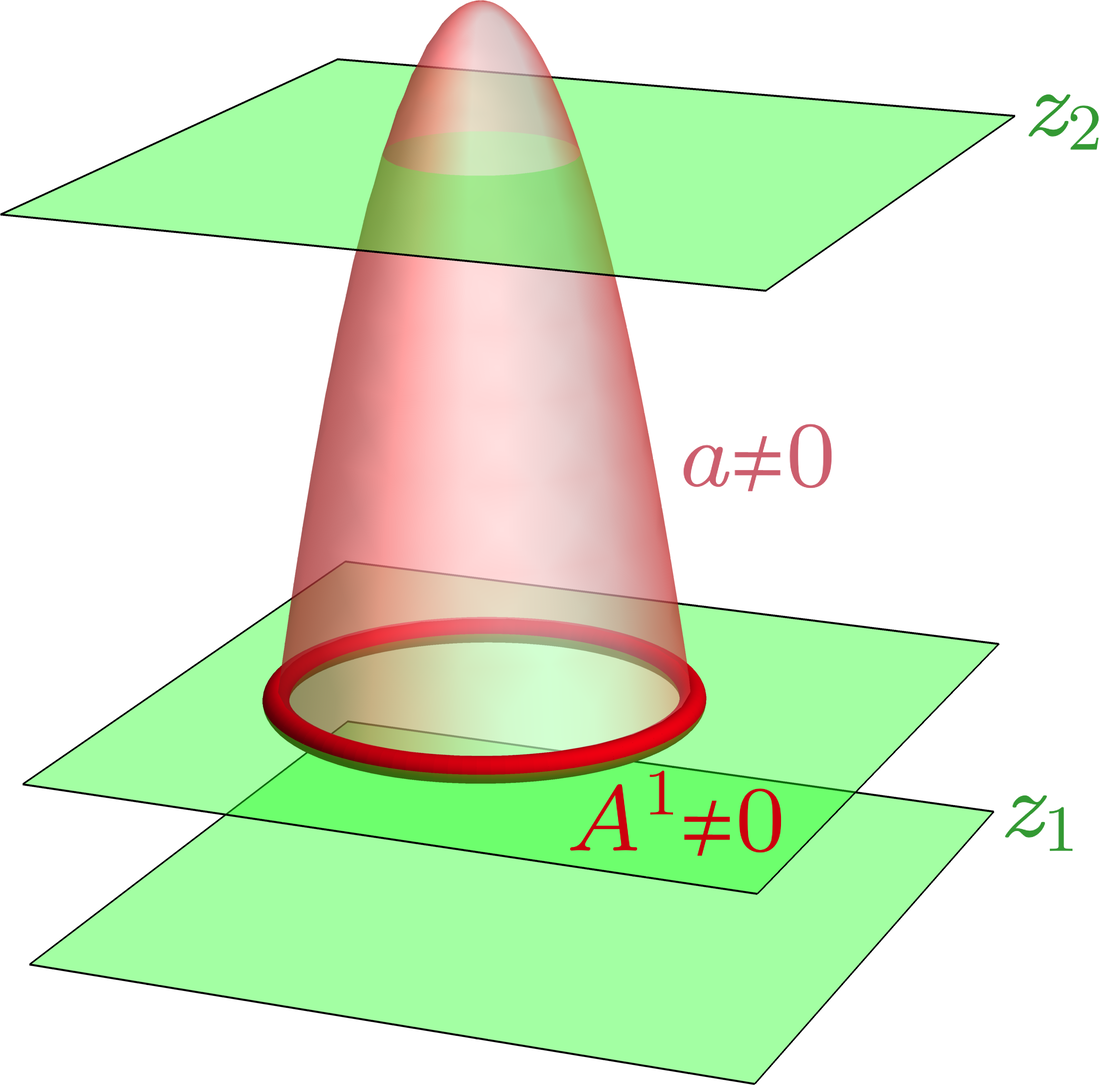}}
  \hspace{.02\columnwidth} \raisebox{1.5cm}{\scalebox{1.7}{$\leftrightarrow$}} \hspace{.02\columnwidth}
  \subfloat[]{\includegraphics[width=.35\columnwidth]{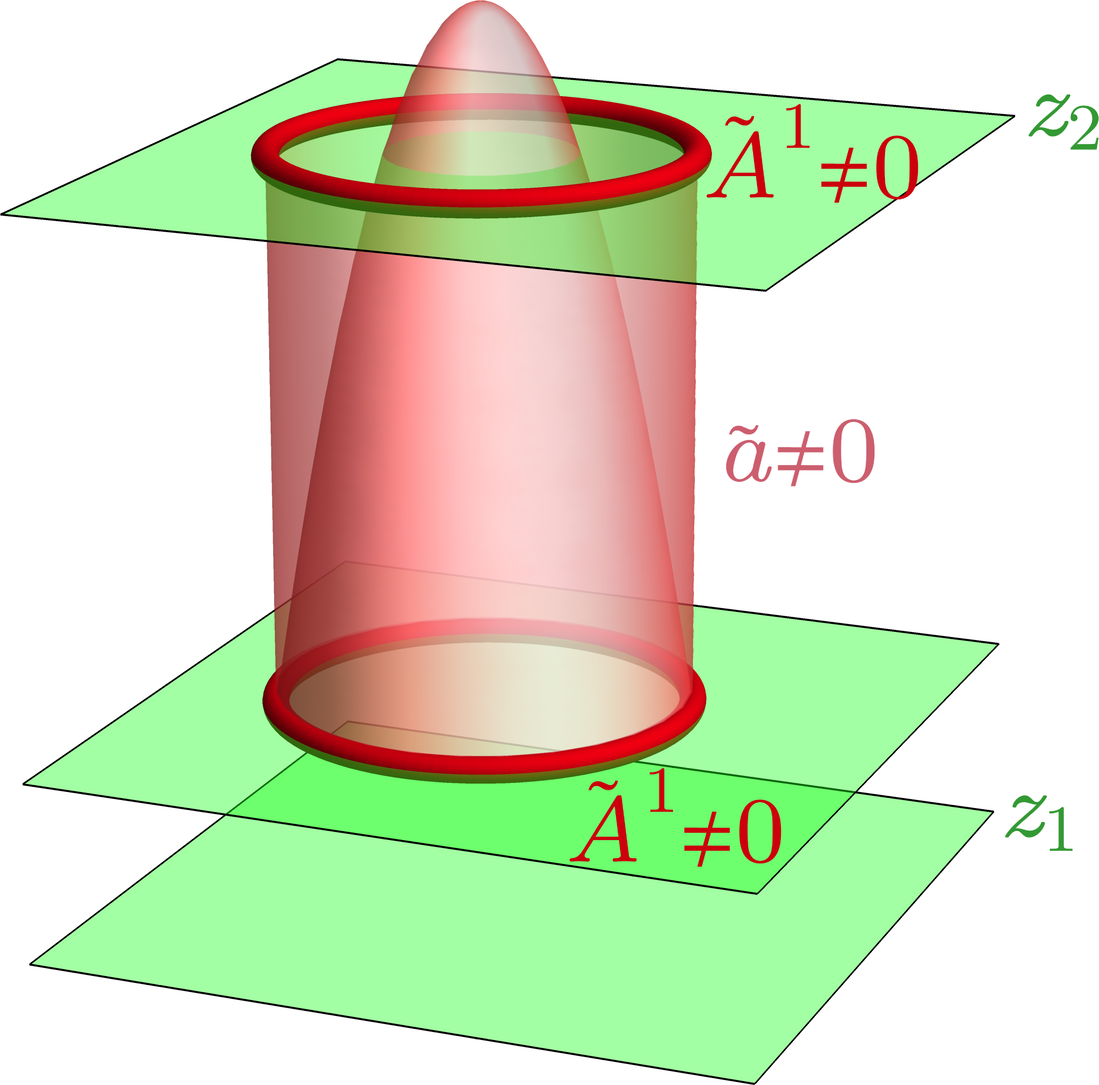}} \\
  \subfloat[]{\includegraphics[width=.35\columnwidth]{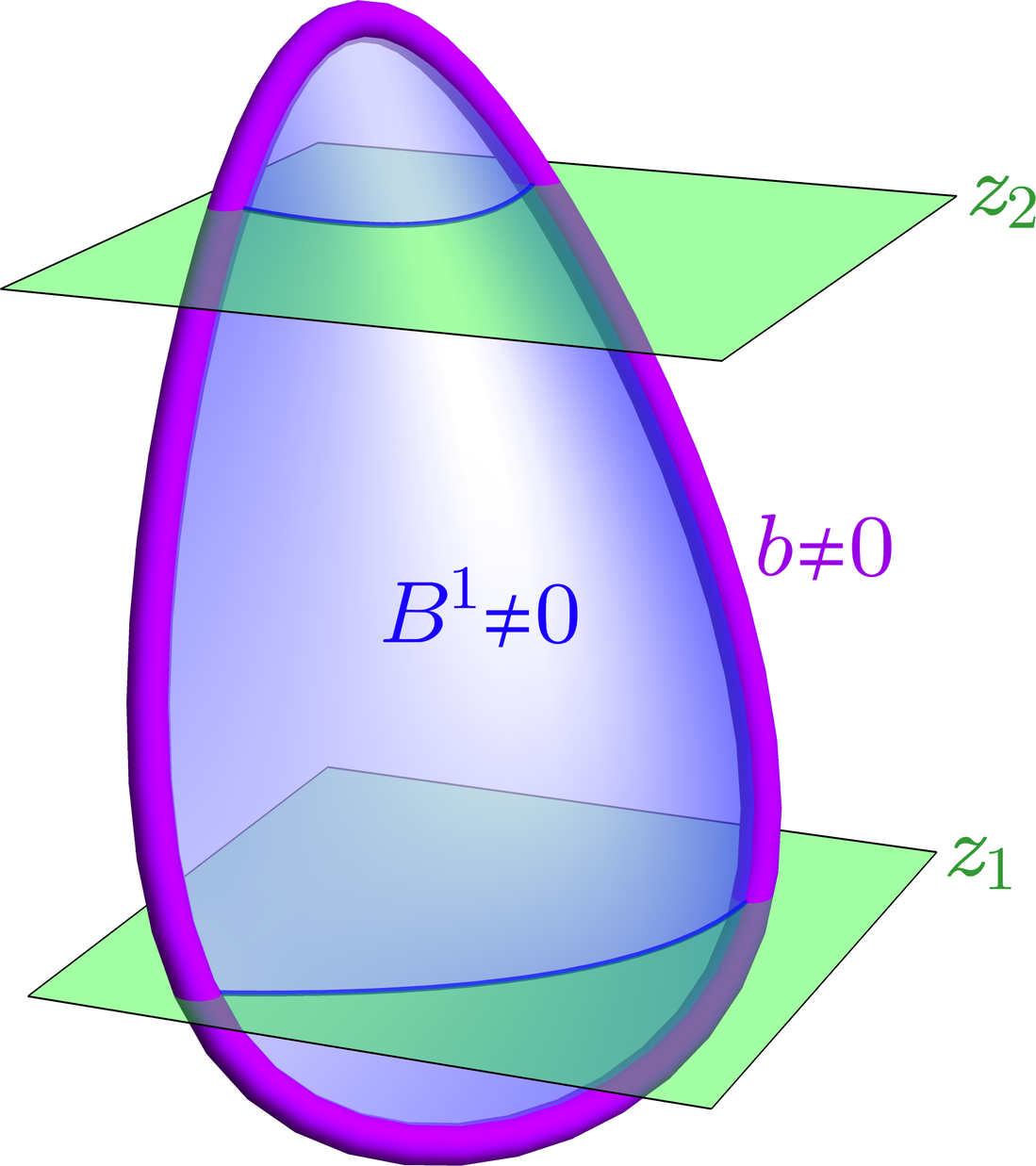}}
  \hspace{.02\columnwidth} \raisebox{1.5cm}{\scalebox{1.7}{$\leftrightarrow$}} \hspace{.02\columnwidth}
  \subfloat[]{\includegraphics[width=.35\columnwidth]{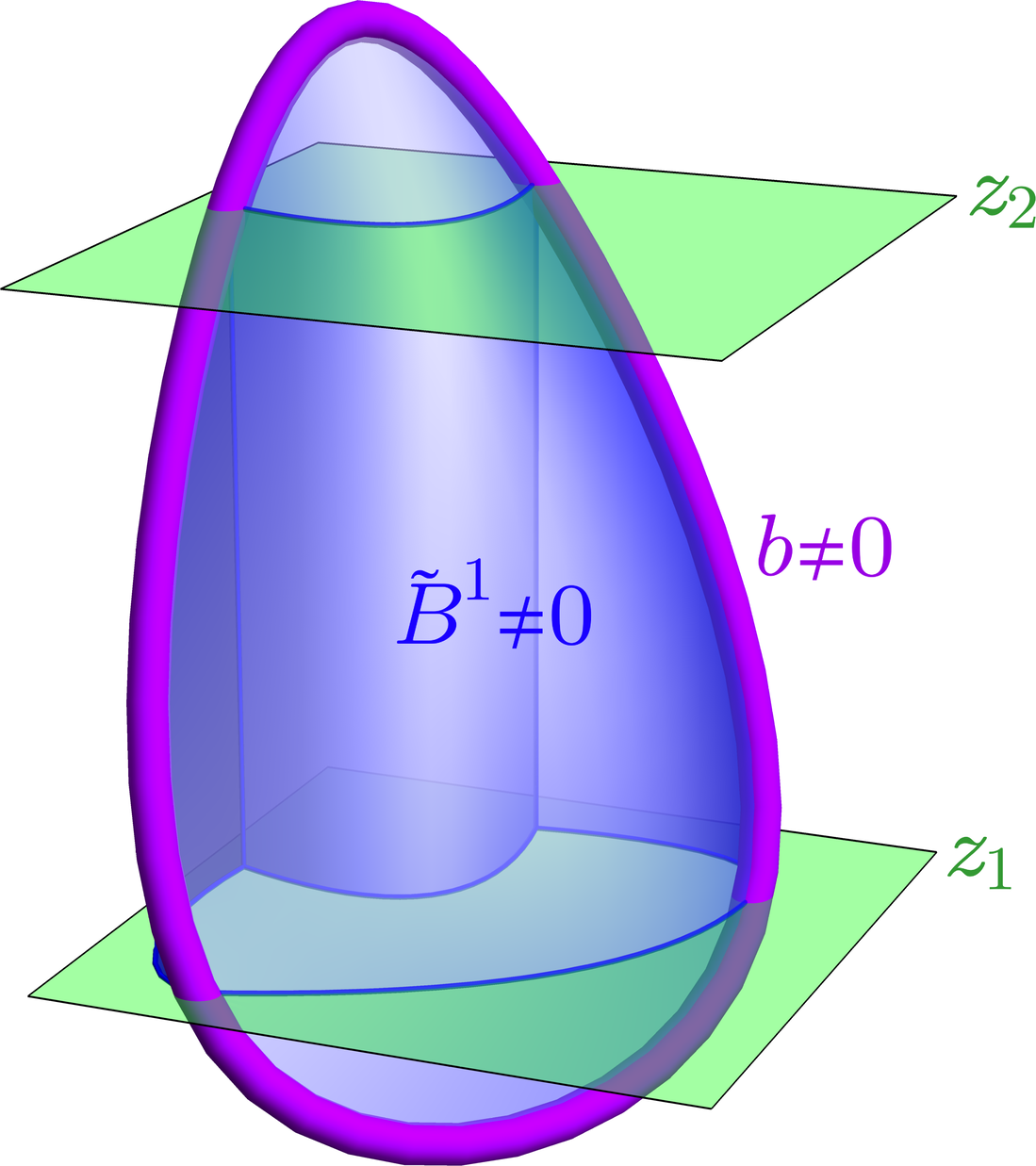}}
  \caption{%
    Examples of how the duality \eqnref{eq:aB duality} acts on field configurations that satisfy the equations of motion
      $da + \sum_k m_k A^k = dA^k = db = (dB^k + n_k b) \wedge e^k = 0$.
    Some leaves are shown in green.
    {\bf (a)} A membrane of $a \neq 0$ (light red) that ends on
      a loop of $A^1 \neq 0$ (red) gets mapped to
    {\bf (b)} the same fields but with the membrane extended vertically (from the $A^1 \neq 0$ loop)
      to $z_2$, ending on a new loop of $\tilde{A}^1 \neq 0$.
    $A^k = 0$ and $a = 0$ elsewhere.
    {\bf (c)} A membrane of $B^1 \neq 0$ (light blue) with a boundary on a loop of $b \neq 0$ (purple).
    (c) is mapped to {\bf (d)}, which adds an additional membrane of $\tilde{B}^1 \neq 0$ within $z_1 < z < z_2$.
    Importantly, note that $da + \sum_k m_k A^k = (dB^1 + n_1 b) \wedge e^1 =0$
      is satisfied everywhere in (a) and (c)
      since membranes of $a \neq 0$ and $B^1 \neq 0$ end on loops of $A^1 \neq 0$ and $B^1 \neq 0$;
      while $d\tilde{a} = d\tilde{B}^1 \wedge e^1 = 0$ is satisfied for $z_1 < z < z_2$ in (b) and (d) since membranes of $\tilde{a} \neq 0$ and $\tilde{B}^1 \neq 0$ do not have boundaries in this region.
  }\label{fig:duality}
\end{figure}

\refcite{3manifolds} showed that a finite-depth local unitary transformation can be used to
  decouple a layer of toric code from the X-cube model and reduce the lattice length of the X-cube model by 1 in one direction [\eqnref{eq:Xcube RG}].
By repeating this process, a stack of many neighboring layers can be exfoliated.
This is the analog of the field theory duality that we study here.

On a lattice, removing $L$ neighboring layers requires a local unitary transformation of depth $O(\log\,L)$
  [by exfoliating every other layer for $O(\log\,L)$ steps].
As such, removing many layers can not be done using a constant-depth local unitary transformation.
The fact that the unitary transformation can not be of constant depth implies that the duality
  maps some local operators to nonlocal operators.

This nonlocality is made explicit by the integrals in \eqnref{eq:aB duality}.
See also \figref{fig:duality},
  which shows an example of how the duality acts on an example field configuration.

The duality can also change the spacetime dimensionality of gauge invariant operators.
For example, if $M_k = N$, $n_k = 1$, and $e^1 = dz$ is the only foliation, then the duality will map:
\begin{equation}
  e^{\ii \oint_{\partial\M_2^\text{F}} a \,+\, \ii \int_{\M_2^\text{F}} A^1} \leftrightarrow e^{\ii \oint_{\partial\M_2^\text{F}} \tilde{a}}
\end{equation}
where $\M_2^\text{F}$ is a disk within the XZ plane and between $z_1 < z < z_2$,
  and $\partial\M_2^\text{F}$ denotes its circular boundary.
The left hand side is a gauge invariant 2D membrane operator,
  which has higher dimension than the 1D string operator on the right hand side.
The right hand side is an example of the string operator in \eqnref{eq:fracton}
  (which is gauge invariant since $\tilde{n}_k = 0$).

Applying the duality throughout the entire spacetime and to all foliations would appear to result in a duality
  to an FQFT with $n_k=0$, which describes decoupled 3+1D and a foliation of 2+1D BF theories.
However, such a mapping would require taking the limit $z_1 \to -\infty$ and $z_2 \to +\infty$ in \eqnref{eq:aB duality},
  which would merely push the nontrivial fracton physics out to infinity.
Nevertheless, between $z_1$ and $z_2$,
  the theory would be a TQFT with no apparent fracton physics remaining in this region.

\subsection{Details}

Below, we show that the duality transformation in \eqnref{eq:aB duality}
  transforms the 
  equations of motion \eqref{eq:EoM J}--\eqref{eq:EoM i} by $n_1 \leftrightarrow \tilde{n}_1(z)$ [\eqnref{eq:n duality}].
(We will assume that there are no source terms: $J^k = 0$, $I^k = 0$, $j=0$, $i=0$.)
\eqsref{eq:EoM I} and \eqref{eq:EoM j} do not transform for $z \neq z_2$ since $n_k$, $a$, and $B^k$ do not appear in these equations of motion.

When $z_1 < z < z_2$, \eqnref{eq:EoM J} transforms as follows:
\begin{align}
  &\quad\; \left\{ dB^1 + n_1 b \right\} \wedge e^1 \\
  &= \left\{ d\tilde{B}^1 + dB^1(z_2) - n_1 d \int_z^{z_2} b + n_1 b \right\} \wedge e^1 \label{eq:J2}\\
  &= \left\{ d\tilde{B}^1 + dB^1(z_2) - n_1 \int_z^{z_2} (db - \dd z \wedge \partial_z b) + n_1 b \right\} \wedge e^1 \label{eq:J3}\\
  &= \left\{ d\tilde{B}^1 + dB^1(z_2) + n_1 [b(z_2) - b] + n_1 b \right\} \wedge e^1 \label{eq:J4}\\
  &= d\tilde{B}^1 \wedge e^1 \label{eq:J5}
\end{align}
\eqnref{eq:J2} results from solving for $B^1$ in \eqnref{eq:aB duality} and plugging that in.
\eqnref{eq:J3} follows from splitting the exterior derivative into spacetime components:
  $\left[ d \int_z^{z_2} b \right] \wedge e^1
    = \sum_{\mu=0,1,2} \dd x^\mu \wedge \left[ \int_z^{z_2} \partial_\mu b \right] \wedge \dd z
    = \left[ \int_z^{z_2} (db - \dd z \wedge \partial_z b) \right] \wedge e^1$.
\eqnref{eq:J4} follows from the equation of motion $db=0$
  and integrating the total derivative $\partial_z b$.
\eqnref{eq:J5} follows from the original equation of motion \eqref{eq:EoM J}: $dB^k + n_k b=0$.

When $z_1 < z < z_2$, \eqnref{eq:EoM i} transforms as follows:
\begin{align}
  &\quad\; da + \sum_k m_k A^k \\
  &= d\tilde{a} - m_1 d\int_{z_1}^z  A^1 + \sum_k          m_k A^k \label{eq:i2}\\
  &= d\tilde{a} - m_1  \int_{z_1}^z dA^1 + \sum_{k \neq 1} m_k A^k \label{eq:i3}\\
  &= d\tilde{a}                          + \sum_{k \neq 1} m_k A^k \label{eq:i4}
\end{align}
\eqnref{eq:i2} results from solving for $a_\mu$ in \eqnref{eq:aB duality} and plugging that in.
\eqnref{eq:i3} follows from splitting the exterior derivative into spacetime components:
  $d\int_{z_1}^z A^1 = \dd x^\mu \wedge \partial_\mu \int_{z_1}^z A^1
    = \sum_{\mu=0,1,2} \int_{z_1}^z \dd x^\mu \wedge \partial_\mu A^1 + \dd z \wedge \partial_z \int_{z_1}^z A^1
    = \int_{z_1}^z d A^1 + A^1$,
  where the last equality makes use of \eqnref{eq:A constraint} and the definition $\big( \int_{z_1}^z A^1 \big)_\mu \equiv \int_{z_1}^z A^1_{3\mu} \, \dd z$ [defined below \eqnref{eq:aB duality}].
\eqnref{eq:i4} follows from the equation of motion \eqref{eq:EoM I}: $dA^k=0$.
The $A^1 \leftrightarrow \tilde{A}^1$ transformation in \eqnref{eq:aB duality} is necessary so that \eqnref{eq:EoM i} remains satisfied at $z=z_2$.



For $z < z_1$ or $z > z_2$,
  the equations of motion do not transforms since
  the duality transformation is trivial in this region of spacetime.
Therefore, we have shown that the duality transformation \eqref{eq:aB duality} transforms the 
  equations of motion \eqref{eq:EoM J}--\eqref{eq:EoM i} by $n_1 \leftrightarrow \tilde{n}_1(z)$ [\eqnref{eq:n duality}].

\section{Connection to Previous Work}
\label{app:previous}

Here, we discuss how the FQFT in \eqnref{eq:L}
  is related to the field theory introduced in \refcite{stringMembraneNet}:
\begin{equation}
  \tilde{L} = \frac{N}{2\pi} \left[ \sum_k B^k \wedge d \tilde{A}^k \wedge e^k + b \wedge d a + \sum_k b \wedge \tilde{A}^k \wedge e^k \right] \label{eq:old L}
\end{equation}
The above Lagrangian is copied from Eq. (3) of \refcite{stringMembraneNet}
  (up to some minus signs),
  except we place a tilde on the 1-form $\tilde{A}^k$ to differentiate it from the foliated (1+1)-form gauge field $A^k$ in this work.

To connect to the FQFT [\eqnref{eq:L}],
  we can make the following hand-wavy replacement in $\tilde{L}$ (and generalize the coefficients):
\begin{equation}
  \tilde{A}^k \wedge e^k \to A^k \label{eq:e absorb}
\end{equation}

As mentioned in Sec. 4.1.1 of \refcite{stringMembraneNet},
  it is difficult to quantize the coefficient $N$ in \eqnref{eq:old L}.
The reason is that rescaling the foliation $e^k \to \gamma e^k$ [\eqnref{eq:e transform}]
  for constant $\gamma$ does not affect the foliation,
  but it would rescale $N$.
By absorbing the foliation field $e^k$ into $\tilde{A}$ as in \eqnref{eq:e absorb},
  the foliation field no longer explicitly appears in the Lagrangian.
This makes it possible to quantize the coefficients of the FQFT.
}

\end{document}